\documentclass[aps,prd,twocolumn,nofootinbib,superscriptaddress,showpacs,floatfix]{revtex4}

\usepackage{graphicx} 
\usepackage{dcolumn}  

\graphicspath{{ps}}

\begin{document}

\title{
 \quad\\[1cm] \Large
 Measurement of the $B^0 -\overline{B}^0$ mixing rate with
 $B^0(\overline{B}^0) \to D^{*\mp} \pi^{\pm}$ partial reconstruction
}

\affiliation{Aomori University, Aomori}
\affiliation{Budker Institute of Nuclear Physics, Novosibirsk}
\affiliation{Chiba University, Chiba}
\affiliation{Chuo University, Tokyo}
\affiliation{University of Cincinnati, Cincinnati, Ohio 45221}
\affiliation{University of Hawaii, Honolulu, Hawaii 96822}
\affiliation{High Energy Accelerator Research Organization (KEK), Tsukuba}
\affiliation{Hiroshima Institute of Technology, Hiroshima}
\affiliation{Institute of High Energy Physics, Chinese Academy of Sciences, Beijing}
\affiliation{Institute of High Energy Physics, Vienna}
\affiliation{Institute for Theoretical and Experimental Physics, Moscow}
\affiliation{J. Stefan Institute, Ljubljana}
\affiliation{Kanagawa University, Yokohama}
\affiliation{Korea University, Seoul}
\affiliation{Kyoto University, Kyoto}
\affiliation{Kyungpook National University, Taegu}
\affiliation{Institut de Physique des Hautes \'Energies, Universit\'e de Lausanne, Lausanne}
\affiliation{University of Ljubljana, Ljubljana}
\affiliation{University of Maribor, Maribor}
\affiliation{University of Melbourne, Victoria}
\affiliation{Nagoya University, Nagoya}
\affiliation{Nara Women's University, Nara}
\affiliation{National Lien-Ho Institute of Technology, Miao Li}
\affiliation{National Taiwan University, Taipei}
\affiliation{H. Niewodniczanski Institute of Nuclear Physics, Krakow}
\affiliation{Nihon Dental College, Niigata}
\affiliation{Niigata University, Niigata}
\affiliation{Osaka City University, Osaka}
\affiliation{Osaka University, Osaka}
\affiliation{Panjab University, Chandigarh}
\affiliation{Peking University, Beijing}
\affiliation{RIKEN BNL Research Center, Upton, New York 11973}
\affiliation{Saga University, Saga}
\affiliation{University of Science and Technology of China, Hefei}
\affiliation{Seoul National University, Seoul}
\affiliation{Sungkyunkwan University, Suwon}
\affiliation{University of Sydney, Sydney NSW}
\affiliation{Tata Institute of Fundamental Research, Bombay}
\affiliation{Toho University, Funabashi}
\affiliation{Tohoku Gakuin University, Tagajo}
\affiliation{Tohoku University, Sendai}
\affiliation{University of Tokyo, Tokyo}
\affiliation{Tokyo Institute of Technology, Tokyo}
\affiliation{Tokyo Metropolitan University, Tokyo}
\affiliation{Tokyo University of Agriculture and Technology, Tokyo}
\affiliation{Toyama National College of Maritime Technology, Toyama}
\affiliation{University of Tsukuba, Tsukuba}
\affiliation{Utkal University, Bhubaneswer}
\affiliation{Virginia Polytechnic Institute and State University, Blacksburg, Virginia 24061}
\affiliation{Yokkaichi University, Yokkaichi}
\affiliation{Yonsei University, Seoul}
  \author{Y.~Zheng}\affiliation{University of Hawaii, Honolulu, Hawaii 96822} 
  \author{T.~E.~Browder}\affiliation{University of Hawaii, Honolulu, Hawaii 96822} 
  \author{K.~Abe}\affiliation{High Energy Accelerator Research Organization (KEK), Tsukuba} 
  \author{K.~Abe}\affiliation{Tohoku Gakuin University, Tagajo} 
  \author{R.~Abe}\affiliation{Niigata University, Niigata} 
  \author{I.~Adachi}\affiliation{High Energy Accelerator Research Organization (KEK), Tsukuba} 
  \author{M.~Akatsu}\affiliation{Nagoya University, Nagoya} 
  \author{Y.~Asano}\affiliation{University of Tsukuba, Tsukuba} 
  \author{T.~Aso}\affiliation{Toyama National College of Maritime Technology, Toyama} 
  \author{T.~Aushev}\affiliation{Institute for Theoretical and Experimental Physics, Moscow} 
% ...
\collaboration{The Belle Collaboration}

 ~~\\

\noaffiliation


\begin{abstract}
We report a measurement of the $B^0-\overline{B}^0$ mixing
parameter $\Delta m_d$  based on a $\rm 29.1~fb^{-1}$ sample of
$\Upsilon (4S)$ resonance decays collected by the Belle detector
at the KEKB asymmetric $e^+ e^-$ collider.  We use events with a
partially reconstructed $B^0(\overline{B}^0) \to D^{*\mp}
\pi^{\pm}$ candidate and where the flavor of the accompanying $B$
meson is identified by the charge of the lepton from a
$B^0(\overline{B}^0) \to X^{\mp} \ell^{\pm} \nu$ decay.  The
proper-time difference between the two $B$ mesons is determined
from the distance between the two decay vertices. From a
simultaneous fit to the proper-time distributions for the
same-flavor ($B^0(\overline{B}^0)$, $\ell^{\pm}$) and
opposite-flavor ($B^0(\overline{B}^0)$, $\ell^{\mp}$) event
samples, we measure the mass difference between the two mass
eigenstates of the neutral $B$ meson to be $\Delta m_d$= $(0.509
\pm 0.017~(\rm stat) \pm 0.020~(\rm syst))~ps^{-1}$.
\end{abstract}
\pacs{11.30.Er, 12.15.Ff, 12.15.Hh, 13.25.Hw \newpage
}

\maketitle


\normalsize

\setcounter{footnote}{0}

\section{INTRODUCTION}
\begin{figure}[!htb]
\begin{center}
    \includegraphics[height=6.cm]{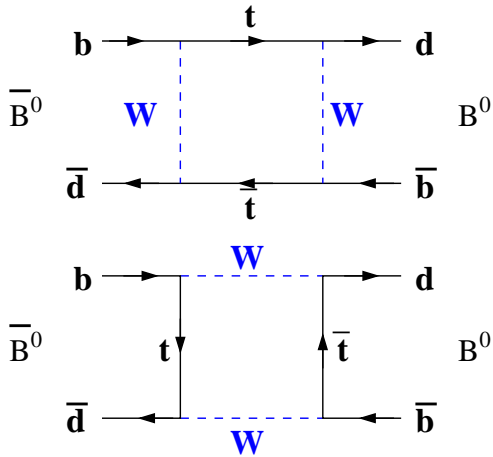} \vskip 4mm
\caption{Standard Model ``box diagrams'' for the second-order weak
 $B^0 -\overline{B}^0$ mixing
 process.}
\label{fig:mixdiag}
\end{center}
\end{figure}

After production, $B^0$ and $\overline{B}^0$ mesons evolve in time
and mix into each other via the second-order weak interaction box
diagrams shown in Fig.~\ref{fig:mixdiag}.
The mixing parameter $\Delta m_d$, which is the mass difference of
the two neutral $B$ mass eigenstates,
is determined from the Feynman diagrams shown in
Fig.~\ref{fig:mixdiag}~\cite{Sandabook} to be
\begin{equation}
\Delta m_d = \frac{G^2_F}{6 \pi^2} f^2_B m_B m^2_W \eta_t S
|V^*_{tb}V_{td}|^2 B_B~,
\end{equation}
where $m_t$, $m_B$ and $m_W$ are the $t$-quark, $B^0$ and $W$
masses; $G_F$ is the Fermi constant; $\eta_t$ is a QCD
correction~\cite{buras1}; $S$ is a function of
${m_t^2}/{m_W^2}$~\cite{Inamiburas2}; $f_B$ is the decay constant
of the $B$ meson; and $B_B$ is the $B$ meson bag
parameter~\cite{donoghue}. In principle, a measurement of the
mixing parameter $\Delta m_d$ can be used to determine the
magnitude of the CKM matrix element $|V_{td}|$. However, there are
large theoretical uncertainties associated with the model
dependence of $f_B$ and $B_B$. We report here a measurement of
$\Delta m_d$ that uses $B$ mesons from the decays of
$\Upsilon(4S)$ states produced by the KEKB collider and recorded
in the Belle detector.  We determine the flavor of one $B$ meson
by partially reconstructing the decays $B^0(\overline{B}^0) \to
D^{*\mp} \pi^{\pm}$; the flavor of the accompanying $B$ meson is
identified by the charge of the lepton from $B^0(\overline{B}^0)
\to X^{\mp} \ell^{\pm} \nu$ decays. In the future, this technique
can be extended to determine the linear combination of CKM angles
$2 \phi_1 + \phi_3$~\cite{dunietz}.

The $\Upsilon(4S)$ decays to a $B^0-\overline{B}^0$ pair that is
nearly at rest in the $\Upsilon(4S)$ center of mass system (CM).
Since $B$ mesons are spin 0 mesons, angular momentum conservation
requires the two $B$ mesons to be in an antisymmetric quantum
state
\begin{equation}
\label{eqn:upsistat} |\alpha \rangle = \frac{1}{\sqrt{2}} \left(
|B^0(1) \rangle|\overline{B}^0(2) \rangle - |\overline{B}^0(1)
\rangle|B^0(2) \rangle \right),
\end{equation}
where 1 and 2 indicate the opposite sides in the $\Upsilon(4S)$
decay plane. The form of Eq.~(\ref{eqn:upsistat}) guarantees that
at any time the amplitude for either $B^0 B^0$ or
$\overline{B}^0\overline{B}^0$ states vanishes. Thus, if we
determine the flavor and decay time $t_{tag}$ for one of the $B$
mesons to decay into a final state $f_{tag}$, then we can measure
the time evolution of the other $B$ at any time $t$ as a function
of the time difference $\Delta t = t - t_{tag}$. For the two-state
neutral $B$ system, the following probability expressions hold to
a good approximation~\cite{Sandabook},
\begin{eqnarray}
\label{eqn:sigpdf} \nonumber \hskip 7mm \matrix{ {\rm P}^{\rm
OF}(\Delta t)&=&{1 \over {4 \tau_{B^0}}}{e^{-{|\Delta t| \over
\tau_{B^0}}}
\left[1+\cos(\Delta m_d \Delta t)\right]~,} \\
{\rm P}^{\rm SF}(\Delta t) &=& { 1 \over {4 \tau_{B^0}}
}{e^{-{|\Delta t| \over \tau_{B^0}}}\left[1-\cos(\Delta m_d \Delta
t)\right]~,}} \hskip -7mm
\end{eqnarray}
where the superscripts SF and OF denote events where the
lepton-tagged and partially reconstructed $B$ mesons have the same
and opposite flavors, respectively.

Since the $B^0$ and $\overline{B}^0$ are nearly at rest in the CM
frame, $\Delta t$ can be determined from the displacement between
the lepton-tagged and partially reconstructed $B$ decay vertices
\begin{equation}
\Delta t \simeq (z_{\rm rec} - z_{\rm tag})/\beta\gamma c
 \equiv \Delta z/\beta\gamma c,
\end{equation}
where the $z$ axis is defined to be anti-parallel to the positron
beam direction and the constant $\beta \gamma = 0.425$ is the
Lorentz boost of the $e^+ e^-$ center of mass system at the KEKB
collider.

This paper is organized as follows: In
Section~\ref{sec:experimental_apparatus}, we describe the KEKB
collider and the Belle detector. The partial reconstruction of
$B^0(\overline{B}^0) \to D^{*\mp} \pi^{\pm}$ decays, the
determination of the $b$-flavor of the accompanying $B$ meson and
the measurement of $\Delta t$ are described in
Section~\ref{sec:evtrecon}. The likelihood fit to the measured
$\Delta t$ distributions is described in Section~\ref{sec:fit}. We
present the results of the fit and studies of sources of
systematic errors in Sections~\ref{sec:fit_results},~\ref{sec:vcheck}
and~\ref{sec:syserr}.  The conclusions are presented in
Section~\ref{sec:conclusion}.

\section{EXPERIMENTAL APPARATUS}
\label{sec:experimental_apparatus}

KEKB~\cite{KEKB} is an asymmetric $e^+e^-$ collider 3~km in
circumference, which consists of 8~GeV $e^-$ and 3.5~GeV $e^+$
storage rings and an injection linear accelerator. It has a single
interaction point (IP) where the $e^+$ and $e^-$ beams collide
with a crossing angle of 22~mrad. The collider has reached a peak
luminosity above 8$\times 10^{33}$~cm$^{-2}$s$^{-1}$. Due to the
energy asymmetry, the $\Upsilon(4S)$ resonance and its daughter
$B$ mesons are produced with a Lorentz boost of $\beta\gamma
=$0.425. On average, the $B$ mesons decay approximately 200~$\mu$m
from the $\Upsilon(4S)$ production point.

The Belle detector~\cite{Belle} is a general-purpose large solid
angle magnetic spectrometer surrounding the interaction point.
Precision tracking and vertex measurements are provided by a
silicon vertex detector (SVD)~\cite{SVD} and a central drift
chamber (CDC)~\cite{CDC} in a 1.5~T magnetic field parallel to the
$z$-axis. The SVD consists of three layers of double-sided silicon
strip detectors (DSSD) arranged in a barrel and covers 86\% of the
solid angle. The three layers at radii of 3.0, 4.5 and 6.0 cm
surround the beam-pipe, a double-wall beryllium cylinder of 2.3~cm
radius and 1~mm  thickness. The strip pitches of each DSSD are
84~$\mu$m for the measurement of the $z$ coordinate and 25~$\mu$m
for the measurement of the $r - \phi$ coordinate. The CDC is a
small-cell cylindrical drift chamber with 50 layers of anode wires
including 18 layers of stereo wires. A low-$Z$ gas mixture
($\mathrm{He}$(50\%)$+ \mathrm{C}_2 \mathrm{H}_6$(50\%)) is used
to minimize multiple Coulomb scattering to ensure good momentum
resolution, especially for low momentum particles. The CDC
provides three-dimensional trajectories of charged particles in
the polar angle region $17^\circ < \theta < 150^\circ$ in the
laboratory frame. The impact parameter resolution for
reconstructed tracks is measured as a function of the track
momentum $p$ (measured in GeV/{\it c}) to be $\sigma_{xy}$ = [19
$\oplus$ 50/($p\beta\sin^{3/2}\theta$)]~$\mu$m and $\sigma_{z}$ =
[36 $\oplus$ 42/($p\beta\sin^{5/2}\theta $)]~$\mu$m. The momentum
resolution of the combined tracking system is $\sigma_{p_t}/p_t =
(0.30/\beta \oplus 0.19p_t)$\%, where $p_t$ is the transverse
momentum in GeV/{\it c}.

The identification of charged pions and kaons uses three detector
systems: the CDC measuring $dE/dx$, a set of time-of-flight
counters (TOF)~\cite{TOF} and a set of aerogel Cherenkov counters
(ACC)~\cite{ACC}. The CDC measures energy loss for charged
particles with a resolution of $\sigma(dE/dx)$ = 6.9\% for
minimum-ionizing pions. The TOF consists of 128 plastic
scintillators viewed on both ends by fine-mesh photo-multipliers
that operate stably in the 1.5~T magnetic field. Their time
resolution is 95~ps ($rms$), providing three standard deviation
 (3$\sigma$) $K^\pm/\pi^\pm$ separation below 1.0~GeV/$c$, and
2$\sigma$ separation up to 1.5~GeV/$c$. The ACC consists of 1188
aerogel blocks with refractive indices between 1.01 and 1.03
depending on the polar angle. Fine-mesh photo-multipliers detect
the Cherenkov light. The effective number of photoelectrons is
approximately 6 for $\beta =1$ particles. Using the information
from these three particle identification systems, the $K /\pi$
likelihood ratio $P(K/\pi) =
{\cal{L}}(K)/({\cal{L}}(K)+{\cal{L}}(\pi))$ is calculated, where
${\cal{L}}(K)$ and ${\cal{L}}(\pi)$ are kaon and pion
likelihoods~\cite{Belle}. A selection with $P(K/\pi)
> 0.6$ retains about 90\% of the charged kaons with a charged pion
misidentification rate of about 6\%.

Photons and other neutral particles are reconstructed in a CsI(Tl)
crystal calorimeter (ECL)~\cite{ECL} consisting of 8736 crystal
blocks, 16.2 radiation lengths ($X_0$) thick. Electron
identification is based on a combination of $dE/dx$ measurements
in the CDC, the response of the ACC, the position and the shape of
the electromagnetic shower, as well as the ratio of the cluster
energy to the particle momentum~\cite{EID}. For the electron
identification requirement used in this analysis, the electron
identification efficiency is determined from two-photon
$e^+e^-\rightarrow e^+e^-e^+e^-$ processes to be more than 90\%
for $p_{\rm lab}>1.0$~GeV/{\it c}. The hadron misidentification
probability, determined using tagged pions from inclusive
$K_S^0\rightarrow \pi^+\pi^-$ decays, is below $0.5\%$.

All the detectors mentioned above are inside a super-conducting
solenoid of 1.7~m radius. The outermost spectrometer subsystem is
a $K_L^0$ and muon detector (KLM)~\cite{KLM}, that consists of 14
layers of iron (4.7~cm thick) absorber alternating with resistive
plate counters (RPC). The KLM system covers polar angles between
20 and 155 degrees. The efficiency of the muon identification
requirement used here, determined by using the two-photon process
$e^+e^-\rightarrow e^+e^-\mu^+\mu^-$ and simulated muons embedded
in $B\overline{B}$ candidate events, is greater than 90\% for
tracks with $p_{\rm lab} > 1$~GeV/{\it c}. The corresponding pion
misidentification probability, determined using $K_S^0\rightarrow
\pi^+\pi^-$ decays, is less than 2\%.

In our analysis, Monte Carlo (MC) events are generated using the
$QQ$ event generator~\cite{QQ} and the response of the Belle
detector is precisely simulated by a GEANT3-based
program~\cite{GEANT}. The simulated events are then reconstructed
and analyzed with the same procedure as is used for the real data.

\section{EVENT RECONSTRUCTION}
\label{sec:evtrecon}

\subsection{Data Sample}
We analyze a $29.1 ~\rm fb^{-1}$ data sample recorded on the
$\Upsilon(4S)$ resonance. The data was taken from June 1999 to
July 2001 and corresponds to about $3.13 \times 10^{7}$
$B\overline{B}$ pairs.

\subsection{$B\overline{B}$ Event Pre-selection}

In Belle, neutral $B$ mesons can only be created via the process
$\Upsilon(4S) \to B^0 \overline{B}^0$. To suppress the non-$b
\overline{b}$ background processes from QED, beam-gas and $e^+ e^-
\to \tau^+ \tau^-$, we select hadronic events using event
multiplicity and total energy variables~\cite{hadsel}.

\subsection{$B \to D^* \pi$ Decay Partial Reconstruction}
\label{sec:partial}

We now describe a partial reconstruction method for the decay
chain $B \to D^* \pi_f$, $D^* \to D \pi_s$, where $\pi_f$ and
$\pi_s$ designate a fast $\pi$ and slow $\pi$,
respectively.\footnote{Throughout the paper, the charge conjugate
process is implied, e.g. $B^0 \to D^{*-} \pi^+$ denotes also
$\overline{B}^0 \to D^{*+} \pi^-$ etc. } This is a variation of a
technique that was first developed by the CLEO
collaboration~\cite{partial}. For the Belle experiment, we modify
and apply this method to make a precise time dependent measurement
of $\Delta m_d$. Unlike analyses that use fully reconstructed $B
\to D^* \pi_f$ decays, we do not use any properties of the decay
products of the $D^0$ meson in the decay $D^* \to D \pi_s$.
According to a Monte Carlo simulation, this method yields an order
of a magnitude more events than a full reconstruction method.

\subsubsection{Kinematics}
We consider five particles in the decay chain: $B, D^*, \pi_f, D$
and $\pi_s$. When reconstructed in this way, the system has $20$
degrees of freedom.

For partial reconstruction, only the $\pi_f$ and $\pi_s$ are used.
The $D$ candidate is not reconstructed.  We can obtain constraints
from 4-momentum conservation for both the decays $B \to D^* \pi_f$
and $D^* \to D \pi_s$ (8 constraints). The $B, ~D^*, ~D, ~\pi_f$
and $\pi_s$ masses from the PDG2000 compilation~\cite{PDG2000}
provide further 5 constraints. In addition, we use the constraint
that the $B$ energy is the CM beam energy of KEKB at the
$\Upsilon(4S)$ divided by 2 (1 constraint). The measurements of
the $\pi_f$ and $\pi_s$ 3-vectors provide 6 constraints. In total
there are 20 constraints, equal to the number of degrees of
freedom of the system. Following previous
analyses~\cite{partial,zyh}, we use two variables to measure the
$B \to D^* \pi$ signal. The first is $D^0$ missing mass,
\begin{eqnarray}
\label{eq:mmdb0} \nonumber M_{D_{miss}}^2 &=& m_B^2 + m_{\pi_f}^2
+ m_{\pi_s}^2 - 2 E_B E_{\pi_f} - 2 E_B E_{\pi_s} \\
&+& 2 E_{\pi_f} E_{\pi_s} + 2 |\vec{p}_B| |\vec{p}_{\pi_f}| \cos
\theta_{B \pi_f} \\
\nonumber &+& 2 |\vec{p}_B| |\vec{p}_{\pi_s} | \cos \theta_{B
\pi_s} - 2 |\vec{p}_{\pi_f}| |\vec{p}_{\pi_s}| \cos \theta_{\pi_f
\pi_s},
\end{eqnarray}
where $E, ~\vec{p},~m$ are the energy, momentum and nominal masses
for the $B,~D^*,~D,~\pi_f$ and $\pi_s$ mesons in the CM frame;
$\theta_{B \pi_{f(s)}}$ is the angle between the directions of
motion of the $B$ and $\pi_{f(s)}$; and $\theta_{\pi_f \pi_s}$ is
the angle between the $\pi_f$ and $\pi_s$. In the CM frame, $\cos
\theta_{B \pi_s} \approx - \cos \theta_{B \pi_f}$; the
analysis~\cite{zyh} shows that this approximation, and the
relation
\begin{equation}
\cos \theta_{B \pi_f} = {{- m_B^2 - m_{\pi_f}^2 + m_{D^*}^2 + 2
E_B E_{\pi_f}} \over {2 |\vec{p}_B| |\vec{p}_{\pi_f}|}},
\label{eq:cosbpif}
\end{equation}
can be used to evaluate $M_{D_{miss}}$ without reconstructing the
$B$ meson's flight direction.  The second variable is the angle
$\theta_{\pi_s}^*$ between the slow pion in the $D^*$ rest frame
and the direction of motion of the $D^*$ in the CM frame. Using
partial reconstruction, it is calculated using the relation,
\begin{equation}
\cos \theta_{\pi_s}^* = {{\beta_{D^*}(E_D^{*} - E_{\pi_s}^{*})}
\over {2 |\vec{p}_{\pi_s}^{\: *}|}} - {{|\vec{p}_D|^{2} -
|\vec{p}_{\pi_s}|^{2}} \over {2 \gamma_{D^*}^2 \beta_{D^*} m_{D^*}
|\vec{p}_{\pi_s}^{\: *}|}}, \label{eq:cospis0}
\end{equation}
where $\gamma_{D^*} \equiv {E_{D^*} / m_{D^*}} = \left(E_B -
\sqrt{|\vec{p}_{\pi_f}|^2 + m_{\pi_f}^2}\right)/m_{D^*}$,
$\beta_{D^*} \equiv \sqrt{1 - \left(1 / \gamma_{D^*}^2 \right)}$
and $|\vec{p}_D| = \sqrt{E_D^2 - m_D^2} = \sqrt{(E_B - E_{\pi_f}
-E_{\pi_s})^2 - m_D^2}$; $E_D^{*}$ and $E_{\pi_s}^{*}$ are the
energy of $D$ and $\pi_s$ in the $D^*$ rest frame;
$\vec{p}_{\pi_s}^{\: *}$ is the momentum of $\pi_s$ in the $D^*$
rest frame. The asterisks denote variables calculated in the $D^*$
rest frame. Calculated in this way, $\cos \theta_{\pi_s}^*$ can
take values outside the physical region due to finite resolutions,
or for background events.

\subsubsection{$B^0(\overline{B}^0) \to D^{*\mp} \pi^{\pm}$ Event Selection}
\label{sec:dpi:sigsel}

The $B^0(\overline{B}^0) \to D^{*\mp} \pi^{\pm}$ candidates are
reconstructed with the following requirements. In the CM frame, we
select $\pi_f$ candidates with momentum $\vec{p}_{\pi_f}$ in the
range $2.05 {\rm ~GeV}/c <|\vec{p}_{\pi_f}|<2.45 {\rm ~GeV}/c$ and
an oppositely charged $\pi_s$ with momentum $|\vec{p}_{\pi_s}|$
less than $0.45 {\rm ~GeV}/c$. We require  $ dr<0.05~ \rm cm$,
$dz<2 \rm ~cm$ for the $\pi_f$ candidate and $dr<0.2~\rm cm$,
$dz<2 \rm ~cm$ for the $\pi_s$ candidate to suppress backgrounds
from beam particles that interact with the residual gas of the
vacuum system or spent-beam particles that strike the vacuum
chamber wall. The variables $dr$ and $dz$ are the distances of
closest approach of the track to the interaction point in the $r -
\phi$ and $z$ planes, respectively. For better vertex resolution,
the SVD is required to provide at least 2 spatial points for a
$\pi_f$ candidate track. The $\pi_f$ candidates are required to
have electron and muon likelihood ratios less than 0.8 to suppress
background from semileptonic $B$ decays. Angular momentum
conservation in the pseudoscalar to vector-pseudoscalar decay $B^0
\to D^{*-} \pi^+$ leads to a distribution proportional to $\cos^2
\theta^*_{\pi_s}$ for the angle $\theta^*_{\pi_s}$. To enhance the
signal to background ratio, we require $0.3<|\cos
\theta^*_{\pi_s}|<1.05$. We then select $B^0 \to D^{*-} \pi^+$
candidates with $D^0$ missing mass $M_{D_{miss}}$ greater than
$1.85~{\rm GeV}/c^2$ and $0.3<|\cos \theta^*_{\pi_s}|<1.05$. These
two requirements define the signal region.
\begin{figure}[!htb]
\begin{center}
\includegraphics[height=7cm]{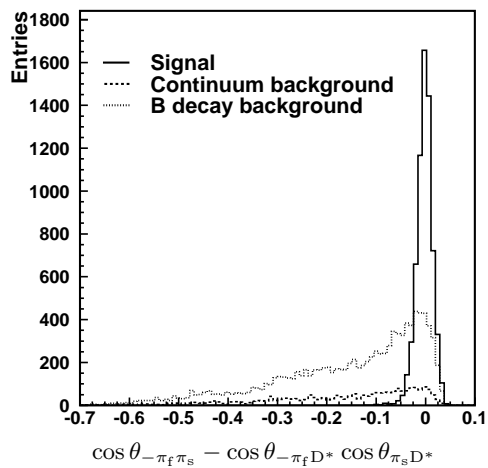}
\\ \vskip -7mm $\rm \cos \theta_{-\pi_f \pi_s} - \cos \theta_{-\pi_f
D^*} \cos \theta_{\pi_s D^*}$ \vskip 7mm

\caption{$\cos \theta_{-\pi_f \pi_s} - \cos \theta_{-\pi_f D^*}
\cos \theta_{\pi_s D^*}$ distributions in a simulation of fully
reconstructed Monte Carlo events.} \label{fig:scos}
\end{center}
\end{figure}

Fig.~\ref{fig:scos} shows that the signal peaks at zero in the
variable $|\cos \theta_{-\pi_f \pi_s} - \cos \theta_{-\pi_f D^*}
\cos \theta_{\pi_s D^*}|$~\cite{zyh}. For about $4\%$ of the
events, more than one pair of opposite sign particles satisfy all
the selection requirements. In these cases, we select the
combination with the smallest value of $|\cos \theta_{-\pi_f
\pi_s} - \cos \theta_{-\pi_f D^*} \cos \theta_{\pi_s D^*}|$.
$\theta_{-\pi_f \pi_s},~\theta_{-\pi_f D^*}$ and $\theta_{\pi_s
D^*}$ are the angles between the directions of $-\vec{p}_{\pi_f}$,
$\vec{p}_{D^*}$ and $\vec{p}_{\pi_s}$ in the $B \to D^* \pi_f$,
$D^* \to D \pi_s$ decay, respectively. This quantity is zero when
the direction of the $\pi_s$ is a good approximation to the $D^*$
direction and is small when the plane defined by the
$\vec{p}_{D^*}$ and $\vec{p}_{\pi_s}$ nearly coincides with the
plane defined by the $\vec{p}_{D^*}$ and $-\vec{p}_{\pi_f}$.

\subsection{Flavor Tagging for neutral $B$ meson}
\label{sec:flavortag}

The flavor of the signal $B$ decay is obtained from the charge of
the fast pion.  The flavor of the accompanying $B$ meson decay is
determined from the charge of the primary lepton ($e$ or $\mu$)
from the semileptonic decay $b\to c \ell^-\nu$. In addition to
tagging the flavor of the $B$ meson decay, there are two other
major reasons for using a high-momentum lepton:  the point of
closest approach of the lepton track and the beam-line provides
the location of the tagging side vertex; the requirement of a high
momentum lepton in the event dramatically reduces the continuum
background.

We require the lepton to have a momentum in the CM frame of at
least ${\rm 1.1~GeV/}c$. We select leptons from well measured
tracks by requiring $dr<0.05 \rm ~cm$, $dz<2 \rm ~cm$. We demand
that the lepton have both $r-\phi$ and $z$ hits in the SVD. To
reject secondary leptons from the decay of the unreconstructed
$D^0$ mesons, we require the cosine of the angle between the
lepton and $\pi_f$ to be greater than $-0.8$. We also reject
lepton tracks that, when combined with any other oppositely
charged track in the event, have an invariant mass that is within
$\rm 50~MeV$ of the $J/\psi$ mass. If more than one lepton in an
event satisfies all of the above criteria, the highest momentum
lepton is selected.

However, not all lepton candidates are primary leptons. According
to a MC study, the background is dominated by three sources. The
first is secondary leptons from charm decays
 (``cascade leptons'') which come from the $b \to c \to s \ell^+
\nu$ decay chain and not directly from $b \to c \ell^-
\overline{\nu}$ like the ``primary leptons''. The charges of the
cascade and primary leptons are opposite and, thus, cascade
leptons can bias the tagged flavor of $B$ mesons and must be well
understood. Two of the requirements described above discriminate
against secondary leptons:  the momentum requirement and the
angular cut with respect to the fast $\pi$ direction. The second
category of incorrect tags is due to leptons from $J / \psi$ and
$\psi(2S)$ decays. Equal numbers of positively and negatively
charged leptons are produced from $J / \psi$ and $\psi(2S)$
decays. Hence, the charge of the observed lepton is uncorrelated
with the $B$ flavor. The third category is composed of hadronic
tracks misidentified as leptons. Particle identification is
applied to suppress this background.

\subsection{Vertex Reconstruction}
\label{sec:vertex} An analysis that relies on $\Delta t$
information requires a measurement of $\Delta t$ in
Eq.~(\ref{eqn:sigpdf}). We use the $z$ difference between $B$
decay vertices in the laboratory system to approximate the
proper-time difference $\Delta t$ by
\begin{eqnarray}
\label{eqn:dz} \Delta z \equiv z_{\pi_f} - z_{l} = \gamma (\Delta
z_{CM} + c \beta \Delta t) \simeq c \beta \gamma \Delta t~,
\end{eqnarray}
where $z_{\pi_f}$ and $z_{l}$ are the $z$ positions of the fast
pion and lepton tracks respectively, and $c$ is the speed of
light. The constant $\beta \gamma = 0.425$ is the Lorentz boost
factor for the $e^+ e^-$ center of mass system in the Belle
experiment. $\Delta z_{CM}$ is the $z$ difference between $B$
decay vertices in the CM frame. Here, we assume that the $B$
mesons are at rest in the CM frame and thus $\Delta z_{CM} \simeq
0$. Therefore we have
\begin{eqnarray}
\Delta t \simeq \frac{\Delta z}{c \beta \gamma}~.
\end{eqnarray}

The $z$ positions are determined from the intersection of the $\rm
\pi_f$ (or lepton track) with the profile of $B$ decay vertices,
which is estimated run-by-run from the profile of the interaction
point ($\sigma^{IP}_x \sim 110~\mu {\rm m},~\sigma^{IP}_y \sim
5~\mu {\rm m},~\sigma^{IP}_z \sim 3500~\mu {\rm m}$) convolved
with the average $B$ flight length ($30~\mu$m in the CM frame).

\subsection{Signal}
Monte Carlo simulation shows that $B^0(\overline{B}^0) \to
D^{*\mp} \rho^{\pm}$ candidates are peaked around $1.865~{\rm
GeV}/c^2$ in the $D^0$ missing mass distribution, the same
location as $B^0(\overline{B}^0) \to D^{*\mp} \pi^{\pm}$ decays.
$B^0(\overline{B}^0) \to D^{*\mp} \rho^{\pm}$ decays can also be
treated as signal for the measurement of $B^0-\overline{B}^0$
mixing. Thus, the signal consists of $B^0(\overline{B}^0) \to
D^{*\mp} \pi^{\pm}$ and $B^0(\overline{B}^0) \to D^{*\mp}
\rho^{\pm}$ decays tagged by primary leptons. Henceforth, for
simplicity we use the notation $B^0(\overline{B}^0) \to D^{*\mp}
h^{\pm}$ to represent both $B^0(\overline{B}^0) \to D^{*\mp}
\pi^{\pm}$ and $B^0(\overline{B}^0) \to D^{*\mp} \rho^{\pm}$
decays.

\subsection{Backgrounds}
\label{sec:bkg} Some other decay modes, such as $B^0 \to D^{**-}
\pi^+$ and $B^0 \to D^{*-} l^+ \nu_l$, also peak in $D^0$ missing
mass, and hence can fake the signal. We therefore divide the
backgrounds into unpeaked and peaked categories. Unpeaked
background is dominated by random combinations of $\pi_f$ and
$\pi_s$ with primary leptons from $B^0$ and $B^{\pm}$ decays, and
combinatorial background from continuum. We verify in MC that the
$\Delta z$ shape of the unpeaked background can be modeled by the
$D^0$ missing mass sideband. The $\Delta z$ shapes of unpeaked
background thus are taken from the $D^0$ missing mass sideband.
The peaked background is dominated by the following sources:
signal $B$ decays ($B^0(\overline{B}^0) \to D^{*\mp} \pi^{\pm}$
and $B^0(\overline{B}^0) \to D^{*\mp} \rho^{\pm}$) with
secondary-lepton tags or fake-lepton tags; $B^0$ $\to$ $D^{**-}$
$\pi^+$, $B^+$ $\to$ $\overline{D}^{**0}$ $\pi^+$ and $B^0$ $\to$
$D^{*-} \pi^+ \pi^0$ decays with primary-lepton tags,
secondary-lepton tags or fake-lepton tags. The $\Delta z$ shapes
of peaked background are determined from Monte Carlo simulation.
The details of the background parameterization are described in
Section~\ref{sec:fit}.

\subsection{Signal Yields}
\begin{figure}[tb]
\begin{center}
\vskip -15mm
\includegraphics[height=8cm]{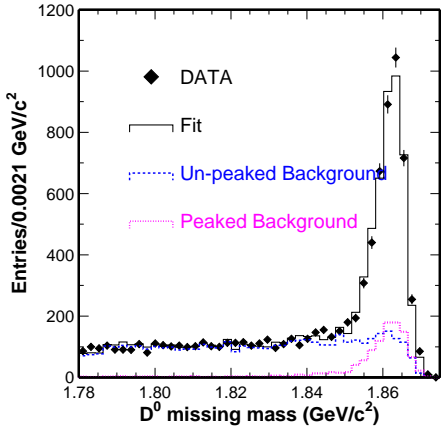}
\caption{$D^0$ missing mass (GeV) distribution for lepton tagged
$B^0 \to D^{*-} h^+$ candidates.} \label{fig:mmdall}
\end{center}
\end{figure}

\begin{figure}[tb]
\begin{center}
\vskip -15mm
    \includegraphics[height=9cm]{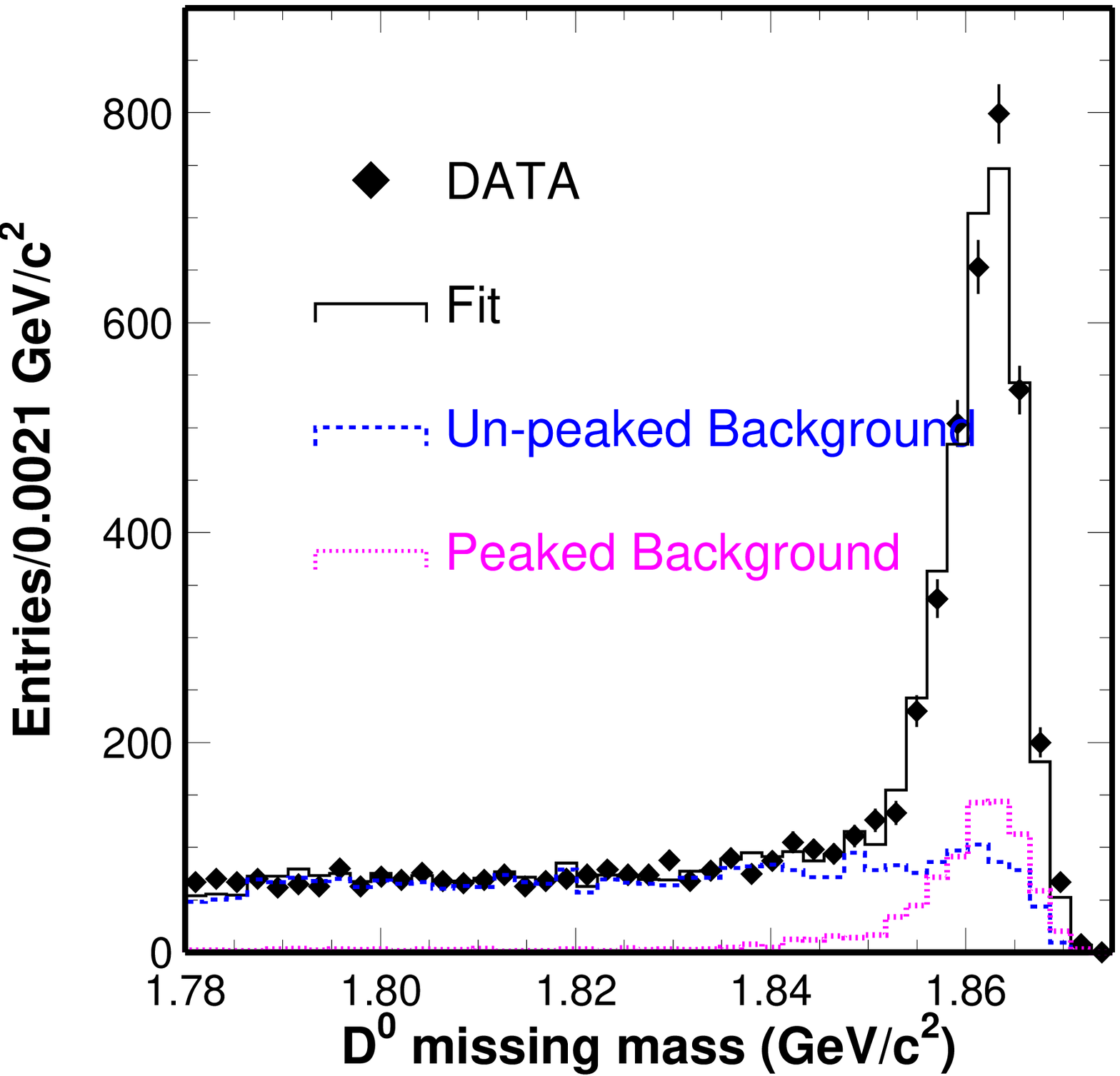} \\
    \vskip -60mm \hskip -35mm (a) \\ \vskip 30mm
    \includegraphics[height=9cm]{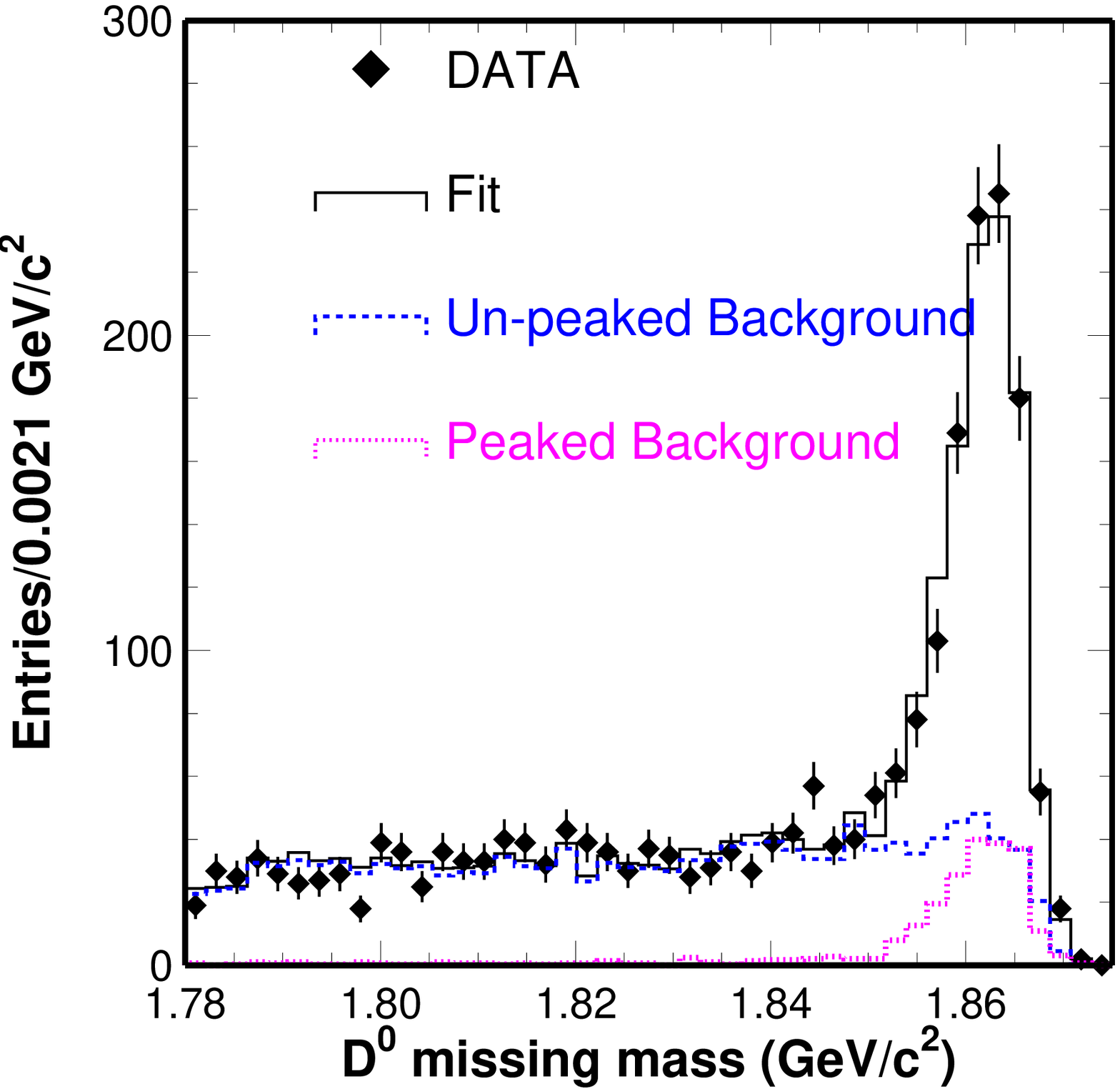} \\
    \vskip -60mm \hskip -35mm (b) \\ \vskip 55mm
\caption{$D^0$ missing mass (${\rm GeV}/c^2$) distribution (a) for
opposite flavor final states and (b) for the same flavor final
states.} \label{fig:mmdosss}
\end{center}
\end{figure}

To obtain the signal yields, we apply a binned maximum likelihood
fit to the $D^0$ meson missing mass distribution. In the fit, the
signal shape is determined from the $B \to D^{*} \pi_f$ Monte
Carlo simulation. The shape of the unpeaked background is taken
from wrong-sign combinations, where the sign of the charges of the
$\pi_f$ and $\pi_s$ are the same. The MC simulation shows that the
wrong sign combinations have a distribution consistent with the
right sign unpeaked background. The peaked background shape comes
from the $B\overline{B}$ Monte Carlo simulation. In the fit, the
shape of the signal, the peaked background and the unpeaked
background are fixed. We also fix the peaked background
normalization from the Monte Carlo simulation. We float the
normalizations of the signal and unpeaked background. By fitting
the $D^0$ missing mass distributions shown in
Figs~\ref{fig:mmdall} and~\ref{fig:mmdosss}, we obtain signal
yields of $3433\pm 81$
 (SF: $751 \pm 41$, OF: $2682 \pm 70$). The estimated backgrounds
 (unpeaked/peaked) in the signal region are $1466 \pm 37$ (SF:
$243 \pm 16$/$219 \pm 15$, OF: $491 \pm 18$/$513 \pm 23$) for
$M_{D_{miss}} > 1.85$ GeV/$c^2$.

\begin{figure}[tb]
\vskip -10mm
\begin{center}
\includegraphics[height=9cm]{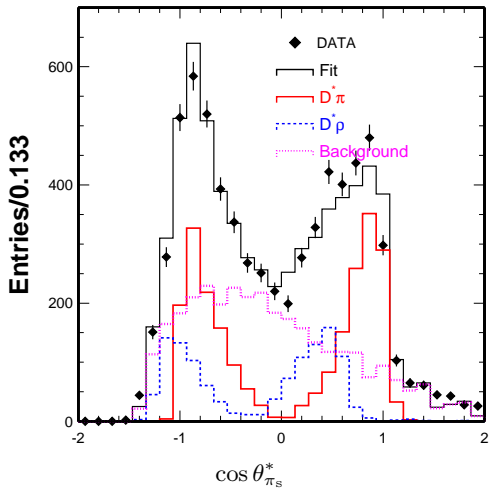}
\\ \vskip -3mm \hskip 0mm $\rm \cos \theta^*_{\pi_s}$ \hskip 0mm
\caption{$\cos \theta^*_{\pi_s}$ distribution. The histogram from
the fit and the contributions of signal and background are
overlaid.} \label{fig:cosall}
\end{center}
\end{figure}

As a consistency check, we use another method to estimate the
signal yields. In Fig.~\ref{fig:cosall}, we fit the $\cos
\theta^*_{\pi_s}$ distribution with both signal and background
shapes determined from MC. Here, we float both the signal and
background normalizations. For the signal shape, we treat $B^0 \to
D^{*-} \pi^+$ and $B^0 \to D^{*-} \rho^+$ events separately. The
polarization of $B^0 \to D^{*-} \rho^+$ is fixed from the CLEO
measurement~\cite{drho}. We obtain $B^0 \to D^{*-} \pi^+$ and $B^0
\to D^{*-} \rho^+$ yields of $2454\pm 90$ and $1353\pm 96$ events.
Note that the yields include incorrectly tagged signal. According
to a MC study, the probability of incorrect tagging is $9.5\%$.
Thus, we conclude that the two methods yield consistent results.
Since the error is smaller for the yields obtained by fitting the
$D^0$ missing mass distribution, we use the results from that
method in the analysis.

\section{MAXIMUM LIKELIHOOD FIT}
\label{sec:fit}

We extract the mixing frequency $\Delta m_d$ by simultaneously
fitting the time evolution distribution of the SF and OF $B^0 \to
D^{*-} h^+$ samples. The unbinned maximum likelihood fitting
method is applied to expressions containing $\Delta m_d$ as a free
parameter, which take into account both signal and background.

\subsection{PDF and Likelihood Function}
\label{sec:pdflike}

Here, we summarize the forms of signal and backgrounds used
in the fitting. The likelihood is also established.

For the signal, the probability density functions (PDF) of OF and
SF events are given by
\begin{eqnarray}
\label{eqn:sigpdf1} F^{\rm OF}_{\rm sig}(\Delta t) &=&
\int^{\infty}_{-\infty} {\rm P}^{\rm OF}(\Delta t')
R_{\rm sig}(\Delta t - \Delta t') d \Delta t' ~,\\
\label{eqn:sigpdf2} F^{\rm SF}_{\rm sig}(\Delta t) &=&
\int^{\infty}_{-\infty} {\rm P}^{\rm SF}(\Delta t') R_{\rm
sig}(\Delta t - \Delta t') d \Delta t' ~,
\end{eqnarray}
where $R_{\rm sig}(\Delta t)$ is the signal resolution function,
which is parameterized by a triple-Gaussian distribution as
follows.
\begin{eqnarray}
\label{eqn:reso} \nonumber R_{\rm sig}(\Delta t) &=& f_1 G(\Delta t; \mu_1,
\sigma_1) + f_2 G(\Delta t; \mu_2, \sigma_2) \\
&&+ (1 - f_1 - f_2) G(\Delta t; \mu_3, \sigma_3)~.
\end{eqnarray}
Here, $G(t; \mu , \sigma)$ is the Gaussian distribution, $f_1$ and
$f_2$ denote the fractions of the first and second Gaussian,
respectively. For the background, the PDFs include separate
contributions for the parts that are peaked and unpeaked in the
$D^0$ missing mass distribution. The time dependent
parameterization includes prompt (zero lifetime) and finite
lifetime components as well as $B^0 - \overline{B}^0$ mixing. The
time distribution of the unpeaked background is determined from
the $D^0$ missing mass sideband data. The functional forms for the
background PDFs are given in the appendix.

Using the PDFs described in Eqs.~(\ref{eqn:sigpdf1})
(\ref{eqn:sigpdf2}) (\ref{eqn:bkgpdf1}) and (\ref{eqn:bkgpdf2}),
the likelihood function can be written as
\begin{eqnarray}
\label{eqn:likli} \nonumber \lefteqn{{\cal L} =  \prod_i
\left((1-f_{\rm bkg})F^{\rm OF}_{\rm sig}(\Delta t_i) + f_{\rm
bkg} f^{\rm
OF}_{\rm bkg}F_{\rm bkg}^{\rm OF}(\Delta t_i) \right) \times } \\
&&\prod_j \left((1-f_{\rm bkg})F^{\rm SF}_{\rm sig}(\Delta t_j)
+ f_{\rm bkg}(1-f_{\rm bkg}^{\rm OF}) F^{\rm SF}_{\rm bkg}(\Delta
t_j) \right)~,
\end{eqnarray}
where $f_{\rm bkg} \equiv N_{\rm bkg} / (N_{\rm bkg} + N_{\rm
sig})$ is the background fraction. $f_{\rm bkg}^{\rm OF}$ is the
fraction of OF events in the background and calculated from
$N_{\rm bkg}^{\rm OF} / N_{\rm bkg}$. The normalizations are
determined from the fit to the $D^0$ missing mass distribution.

\subsection{Resolution Function}
\label{sec:reso} 
\begin{figure}[!htb]
\begin{center}
    \includegraphics[height=7cm]{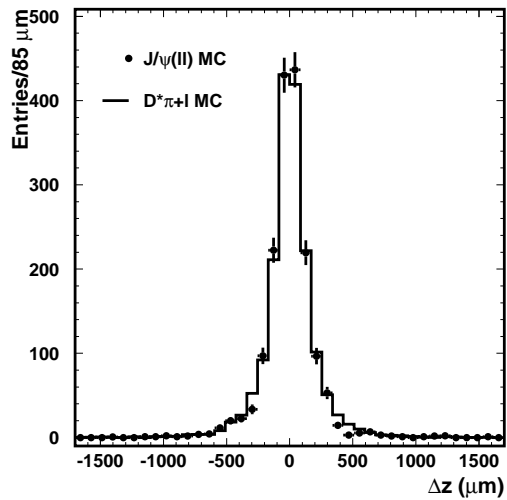}
\caption{Resolution
functions for $J/\psi \to l^+ l^-$ decays and ``$D^{*-}\pi^+ +
l^-_{tag}$'' in MC.} \label{fig:jpsivsreso}
\end{center}
\end{figure}

We need to determine the detector resolution
function to smear the theoretical probability expressions in
Eq.~{\ref{eqn:sigpdf}}. The signal resolution function $R_{\rm
sig}(\Delta t)$ is the distribution of the difference between the
generated and reconstructed $B^0$ decay proper times. A good
approximation to the resolution function can be obtained from the
reconstructed $\Delta z \equiv z_{l^+} - z_{l^-}$ distribution of
$J/\psi \to l^+ l^-$ decays in data, which were parameterized
using the triple-Gaussian distribution in Eq.~(\ref{eqn:reso}). In
Monte Carlo simulation, Fig.~\ref{fig:jpsivsreso} shows good
agreement between the signal resolution function from
``$D^{*-}\pi^+ + l^-_{tag}$'' events and the background subtracted
$\Delta z$ distribution from $J/\psi \to l^+ l^-$ decays. This
indicates that using inclusive $J/\psi$ decays to model the tagged
$D^* \pi$ resolution function is justified.

\begin{figure}[!htb]
\begin{center}
  \hskip -10mm \vskip -10mm
    \includegraphics[height=9cm]{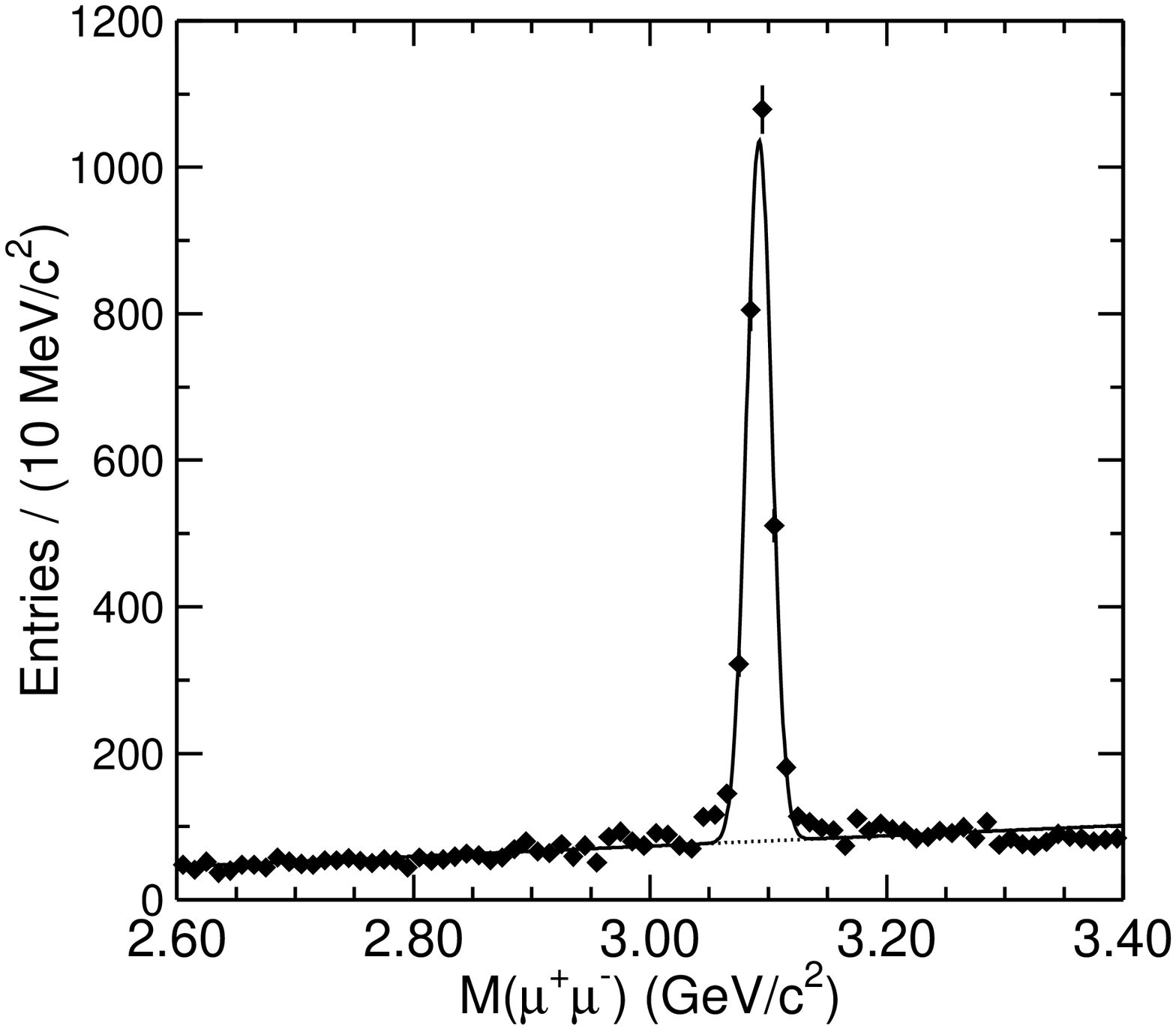}
    \\ \vskip -55mm \hskip -30mm (a) \\ \vskip 15mm
    \includegraphics[height=9cm]{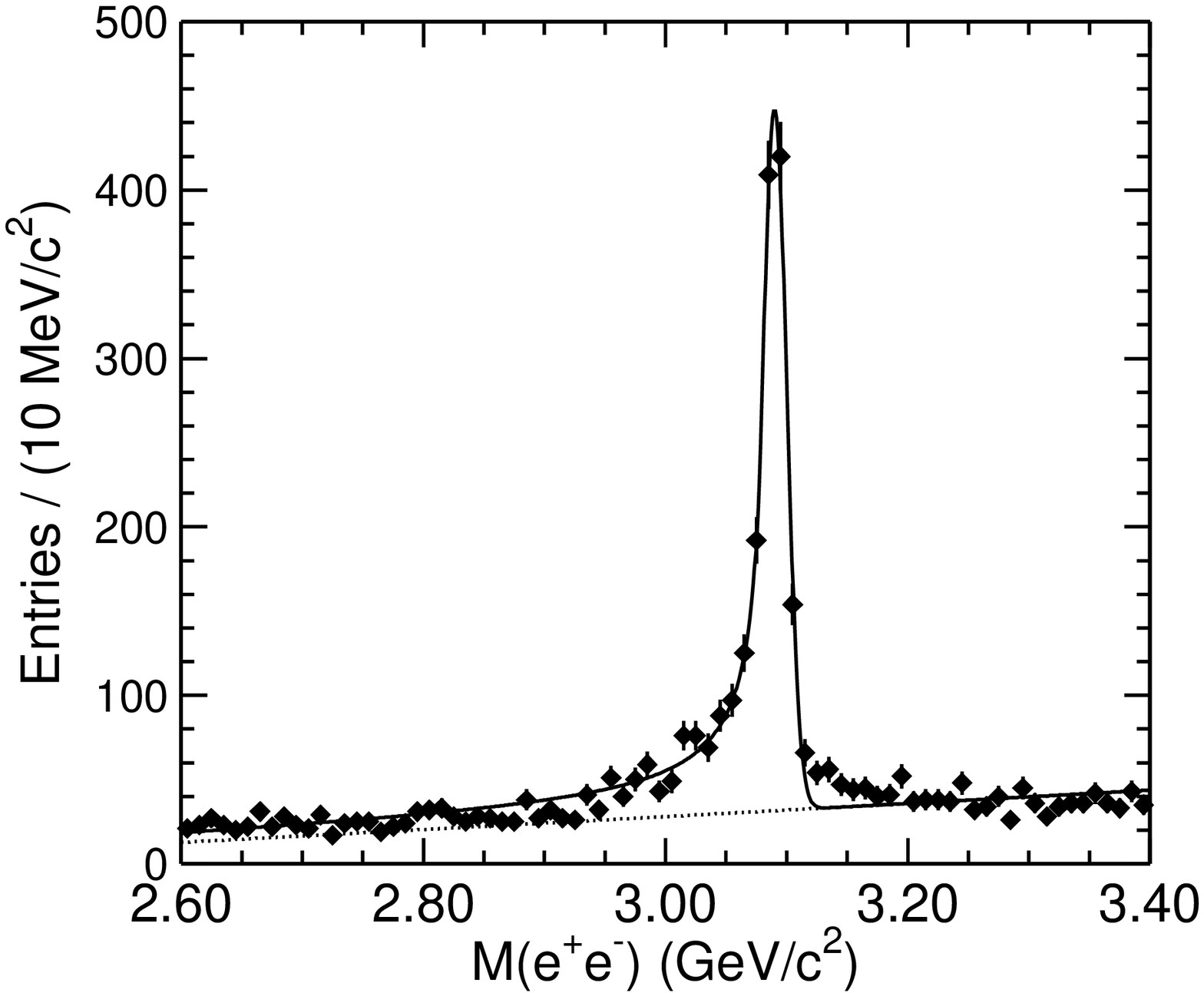} \\
    \vskip -50mm \hskip -30mm (b) \\ \vskip 60mm
\caption{Invariant mass distributions for (a) $J/\psi \to \mu^+
\mu^-$ and (b) $J/\psi \to e^+ e^-$. Fits are described in the
text.} \label{fig:minvjpsi}
\end{center}
\end{figure}

$J/ \psi \to l^+ l^-$ candidates were selected using similar
requirements to those in the ``$B^0 \to D^{*-}\pi^+ + l^-_{tag}$''
selection. The particle identification criteria are changed to
select lepton pairs. Furthermore, no requirement is made on the
angle between the tracks, since $J/\psi$ decays produce lepton
pairs with large opening angles.\footnote{Removing this
requirement is found not to affect the measured resolution
function.} In Fig.~\ref{fig:minvjpsi}, the $J/\psi \to \mu^+
\mu^-$ mass distribution is fitted with the sum of a Gaussian and
a second order polynomial. For the $J/\psi \to e^+ e^-$ case, the
sum of a ``Crystal Ball" function~\cite{cystalballfunc} and a
second order polynomial is used.

By fitting the $J/\psi$ invariant mass distribution in
Fig.~\ref{fig:minvjpsi}, we are able to extract the number of
signal and background events for both the $e^+ e^-$ and $\mu^+
\mu^-$ cases. To obtain all the parameters of the signal
resolution function, we apply a maximum likelihood fit to the
overall $\Delta z$ distribution. In the $\Delta z$ fit, the
background shape is determined from the upper sideband of dilepton
mass.  The normalization is obtained from a second order
polynomial fit to the dilepton mass distribution. The background
shapes are also parameterized by triple-Gaussian functions. The
parameters of the signal resolution function
 (Eq.~(\ref{eqn:reso})) are $\rm \sigma_1 = (94 \pm 6)~\mu m$, $\rm
\sigma_2 = (227\pm 18)~\mu m$, $\rm \sigma_3 = (736 \pm 98)~\mu
m$, $ f_1 = 0.56 \pm 0.07$ and $ f_2 = 0.38 \pm 0.04$. Here, the
fit gives mean values, $\mu_i=0,~i=1,2,3$, for each Gaussian that
are consistent with zero.
\begin{figure}[!htb]
\begin{center}
    \includegraphics[height=7cm]{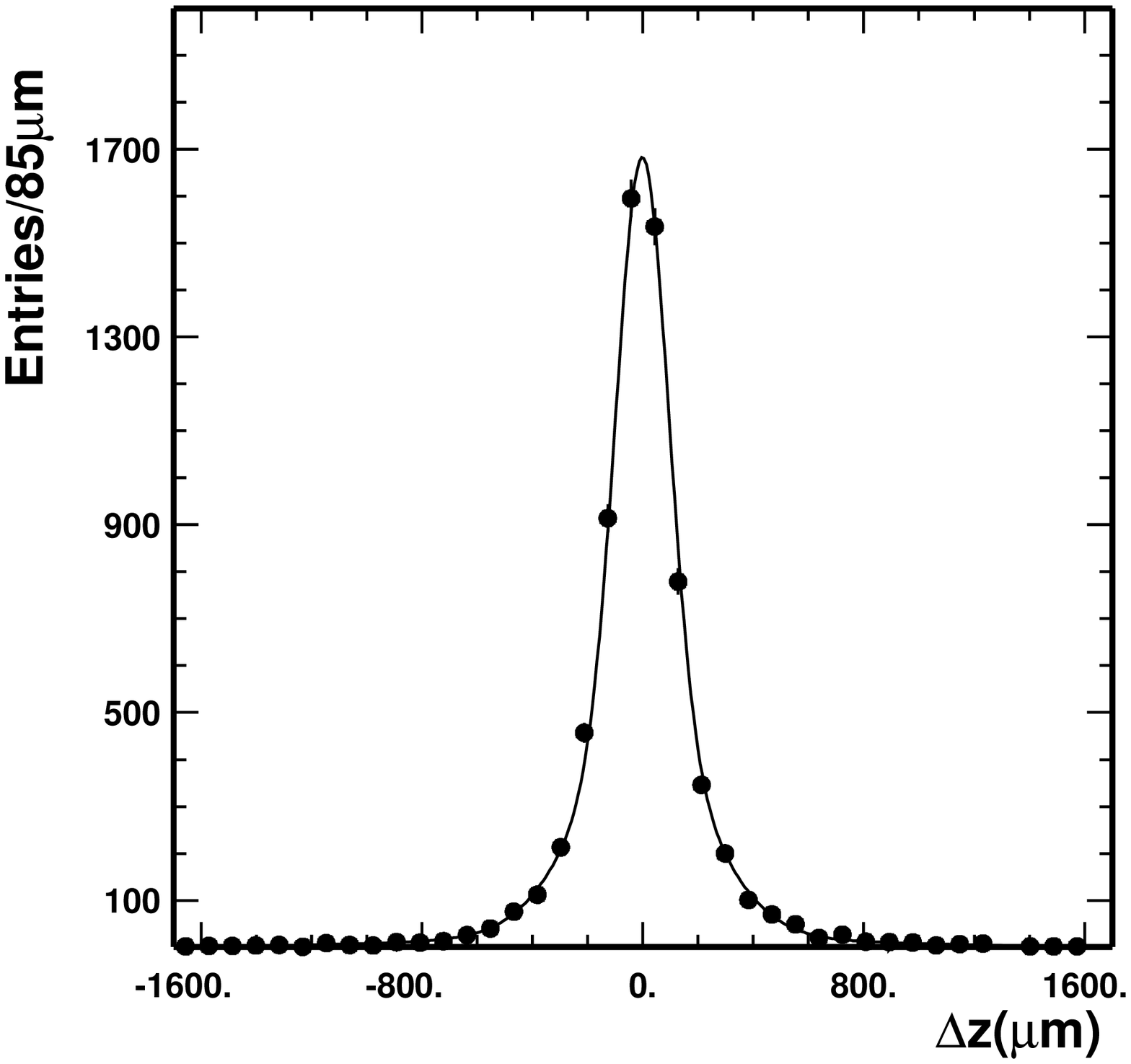}\\ \vskip
    -55mm \hskip -35mm (a) \\ \vskip 40mm
    \includegraphics[height=7cm]{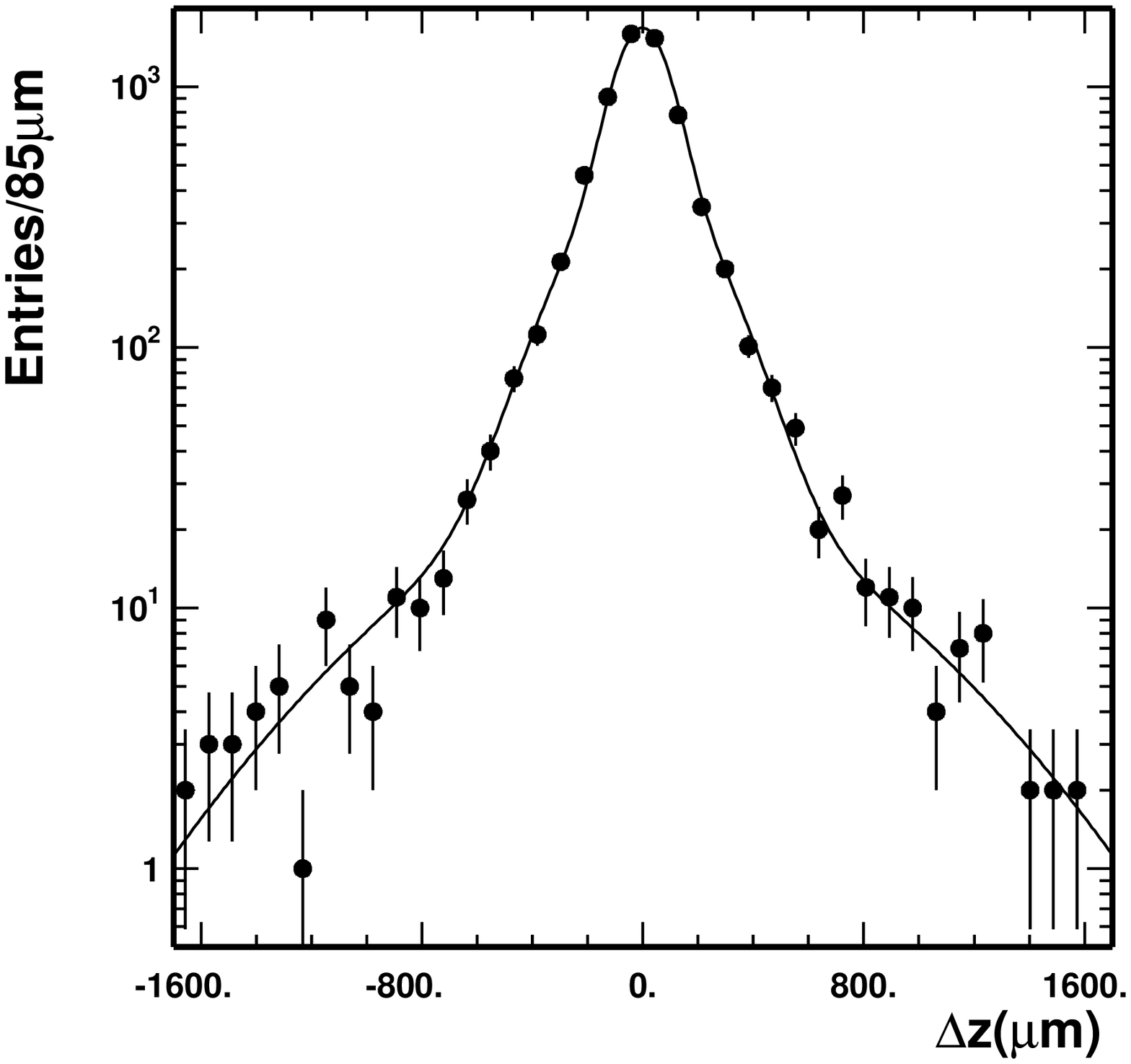} \\ \vskip
-55mm \hskip -35mm (b) \vskip 55mm \caption{Signal resolution
function fit with a triple-Gaussian for $J/\psi \to l^+ l^-$
decays in the data. Plot (a) shows the fit on a linear scale and
plot (b) shows the fit on a semilogarithmic scale.}
\label{fig:reso}
\end{center}
\end{figure}

Similarly, after applying the same technique to MC samples of
$J/\psi \to l^+ l^-$ decays, we find the parameters of the MC
resolution function $R_{\rm MC}(\Delta t)$ to be $\rm \sigma_1 =
(79 \pm 4)~\mu m$, $\rm \sigma_2 = (203\pm 13)~\mu m$, $\rm
\sigma_3 = (625 \pm 74)~\mu m$, $f_1 = 0.49 \pm 0.08$ and $f_2 =
0.43 \pm 0.04$. We also fix the mean values of $\mu_i=0,~i=1,2,3$
to zero.

\begin{figure}[!htb]
\begin{center}
    \includegraphics[height=7cm]{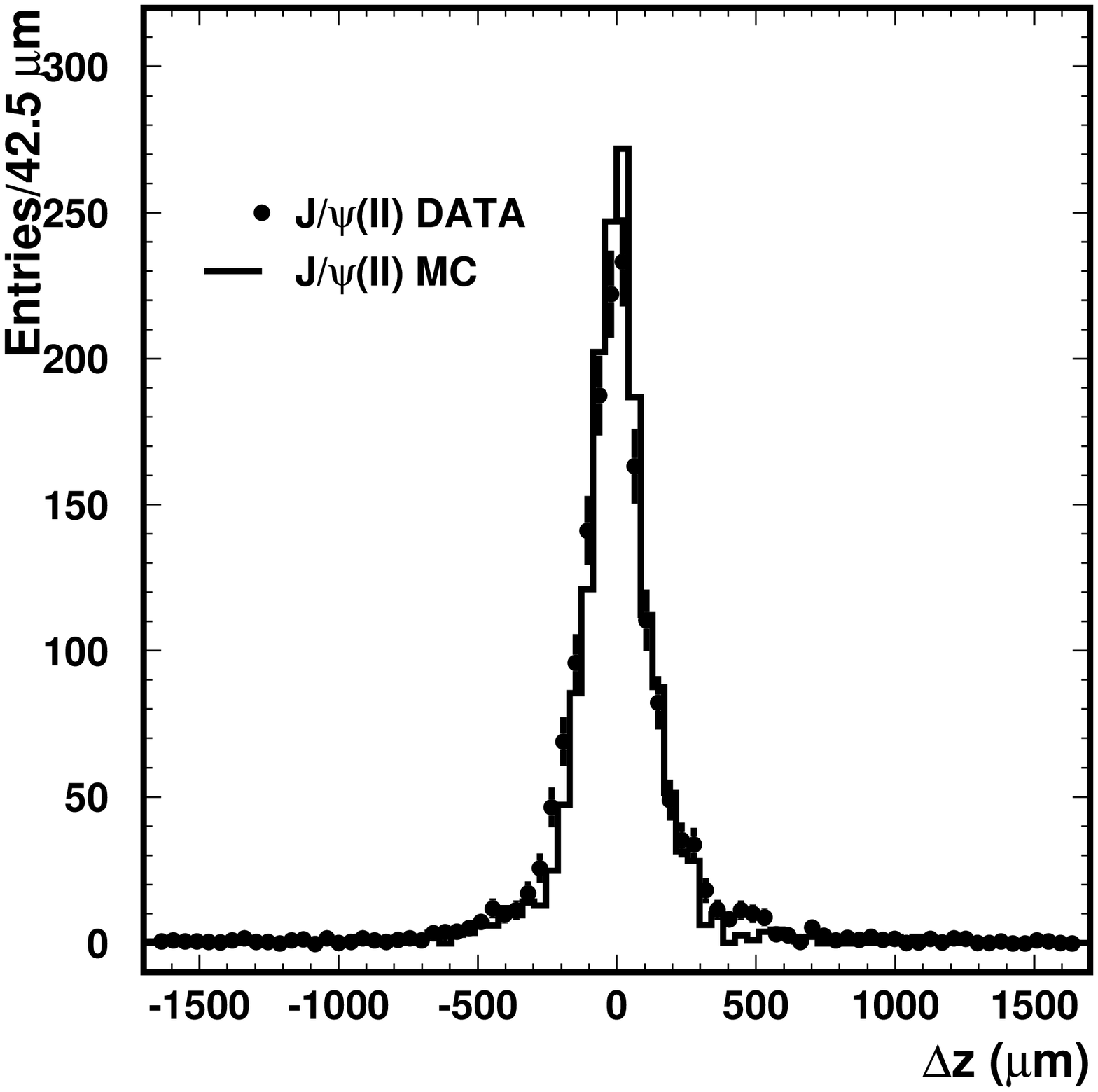}\\
    \vskip -64mm \hskip 40mm (a) \\ \vskip 55mm
    \includegraphics[height=7cm]{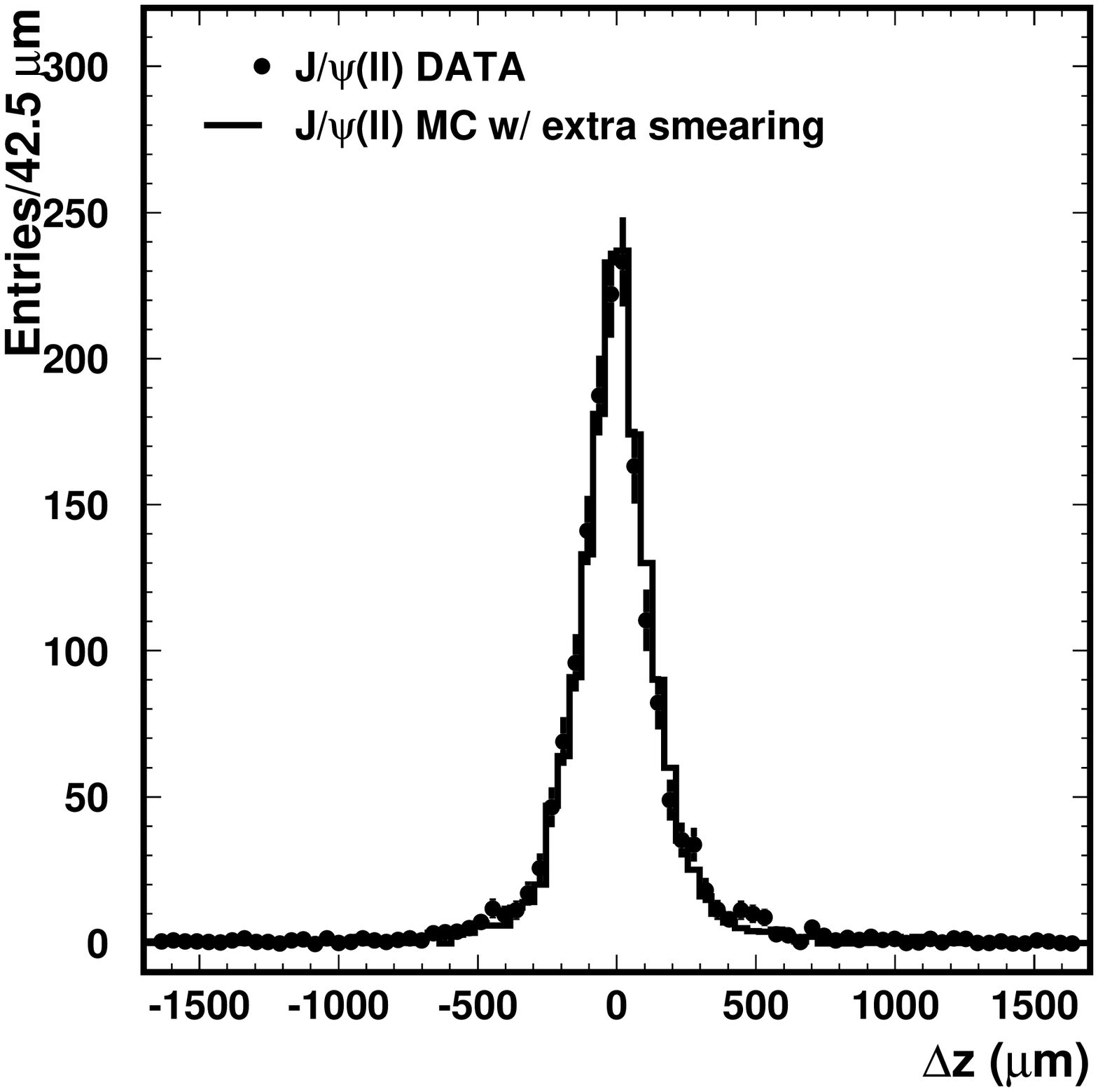} \\
    \vskip -64mm \hskip 40mm (b) \\ \vskip 60mm
\caption{$\Delta z$ distribution for $J/\psi \to l^+ l^-$ decays
in data and Monte Carlo without (a) and with (b) the extra
$50~\mu$m smearing.} \label{fig:smear}
\end{center}
\end{figure}

From the $J/\psi \to l^+ l^-$ sample (see Fig.~\ref{fig:smear}),
we observe that there is a discrepancy between the distributions
in data and MC simulation. To account for this difference, we
convolve the resolution function in MC with a Gaussian of
$50~\mu$m width to reproduce the resolution in the data. This
additional smearing is applied to the MC simulations of peaked
backgrounds. This procedure was verified by comparing the $\Delta
z$ distributions of the $D^0$ missing mass sideband in data and
MC.

For the peaked background case, we use the resolution function for
$J/\psi \to l^+ l^-$ decays in MC with an additional $50~\mu$m
smearing ($R_{\rm bkg}^{\rm peak}(\Delta t) = R_{\rm MC}(\Delta t)
\otimes G(\Delta t; \mu=0, \sigma=50~\mu{\rm m})$). With the
additional $50~\mu$m smearing, we extract the resolution function
of the MC simulation. We find that the mean value of the second
Gaussian of the resolution function for the same flavor peaked
background events is inconsistent with zero, $\mu_2 = (81 \pm 45)
~\mu$m. The corresponding mean value for the opposite flavor
peaked background is consistent with zero. Thus an offset is
included only in the same flavor peaked background shape.

\subsection{Background Distributions}
\label{sec:bkgdis} 
\begin{figure}[!htb]
\begin{center}
    \includegraphics[height=7cm]{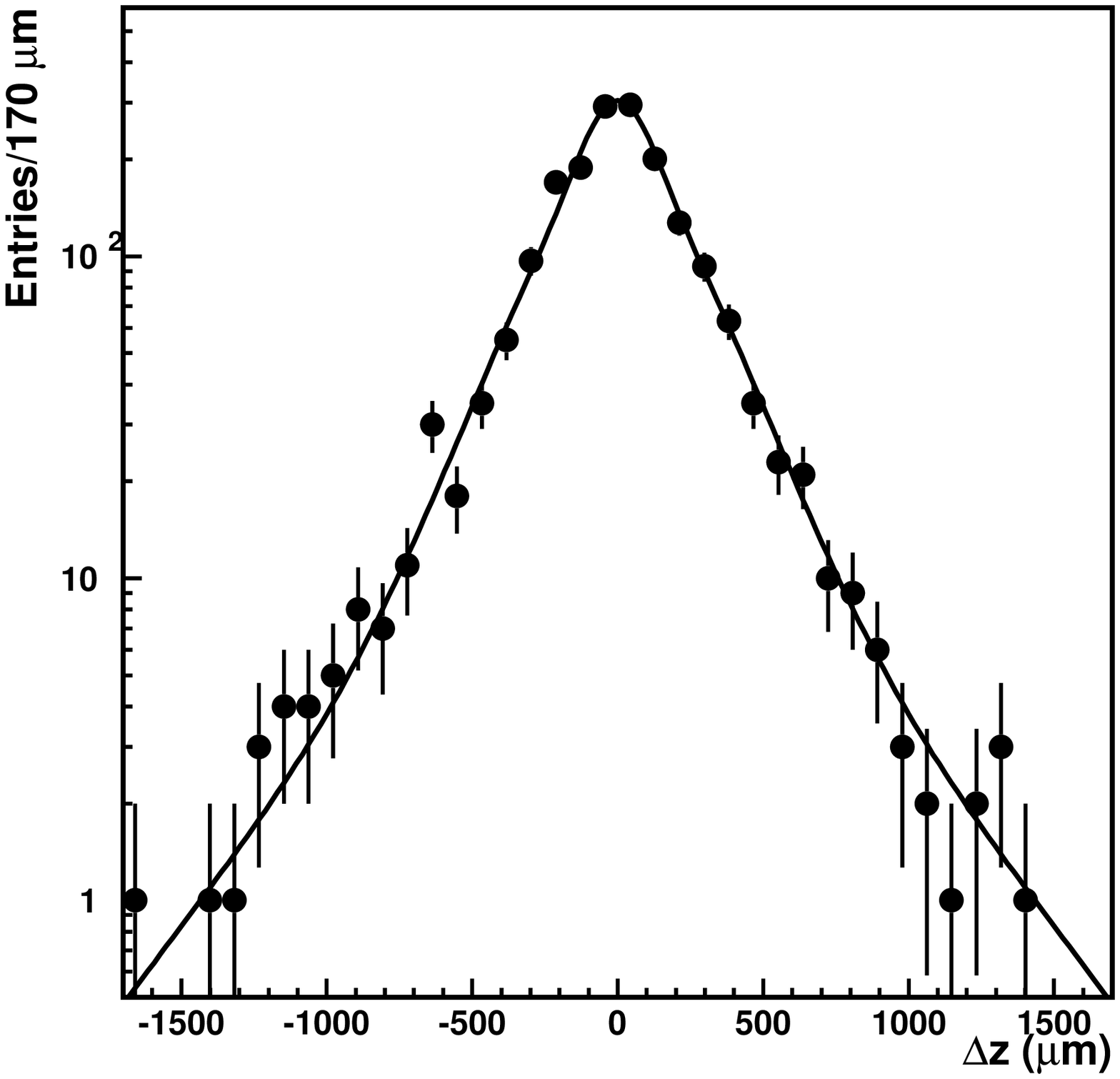}\\
    \vskip -60mm \hskip 40mm (a) \\ \vskip 51mm
    \includegraphics[height=7cm]{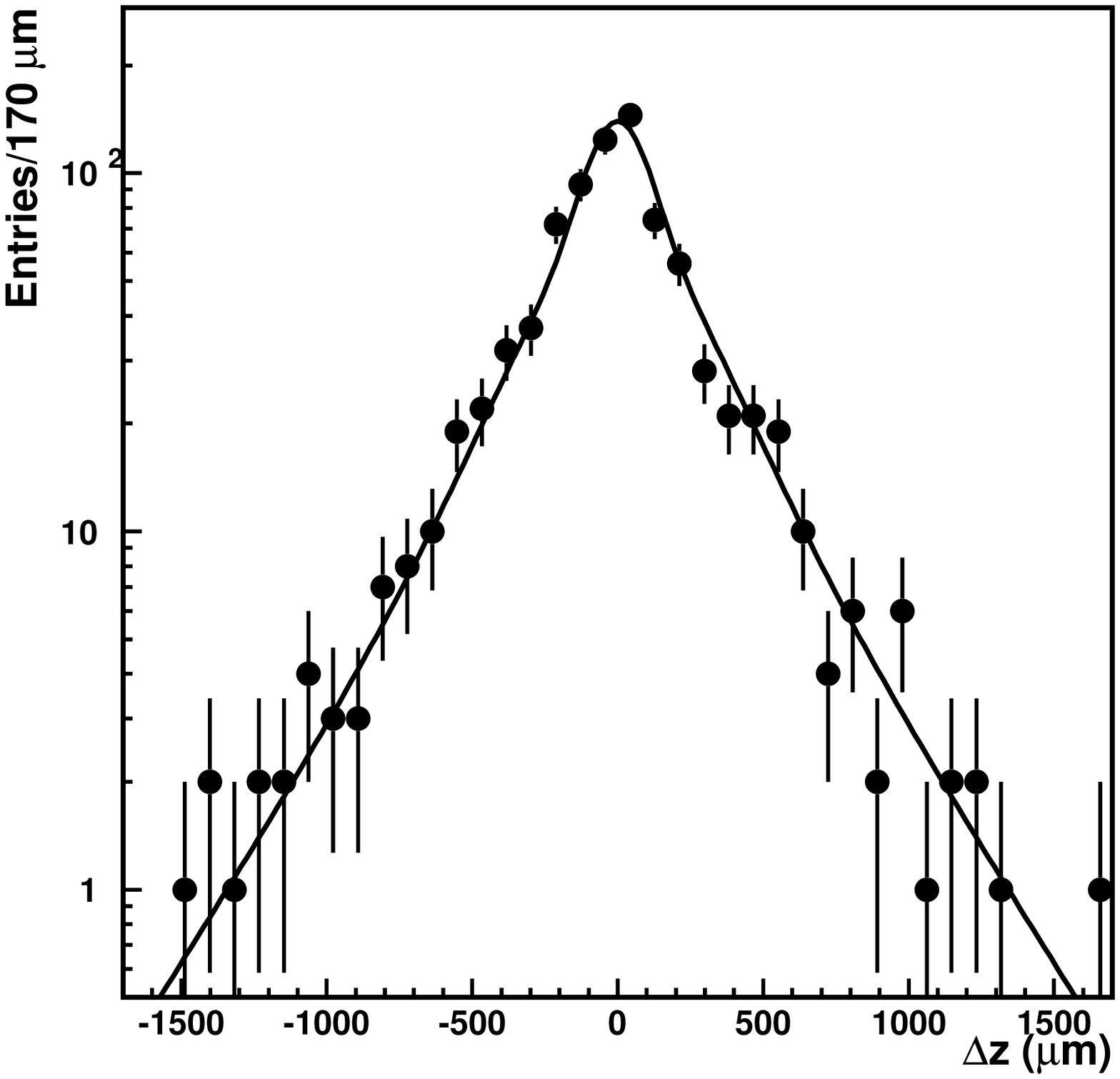} \\
    \vskip -60mm \hskip 40mm (b) \\ \vskip 56mm
\caption{Semilogarithmic plot of the $\Delta z$ distributions from
the $D^0$ missing mass sideband for unpeaked backgrounds in (a)
opposite-flavor and (b) same-flavor events.} \label{fig:bkgunpeak}
\end{center}
\end{figure}

\begin{figure}[tb]
\begin{center}
\vskip -5mm
    \includegraphics[height=7cm]{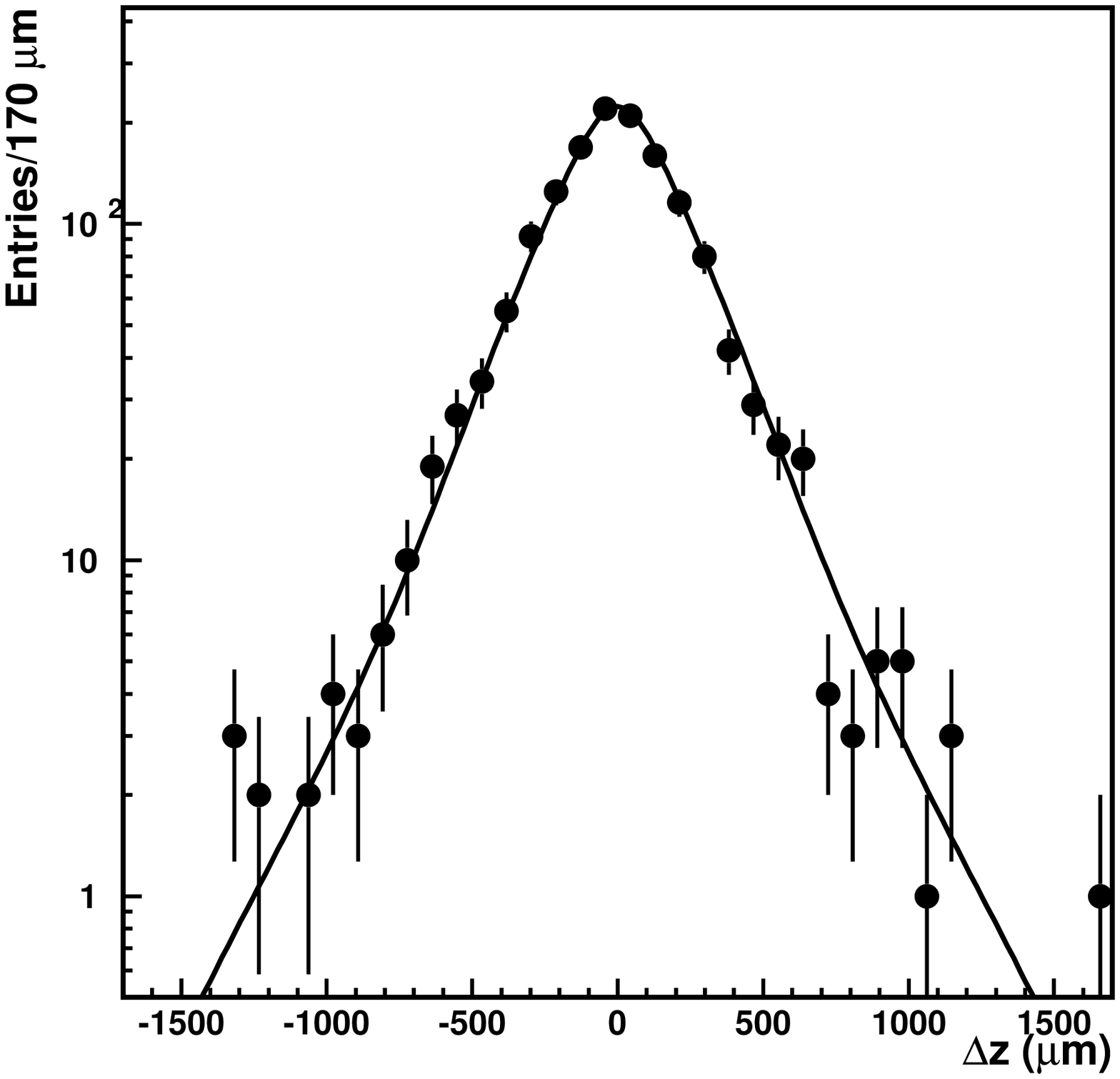}\\
    \vskip -60mm \hskip 45mm (a) \\ \vskip 51mm
    \includegraphics[height=7cm]{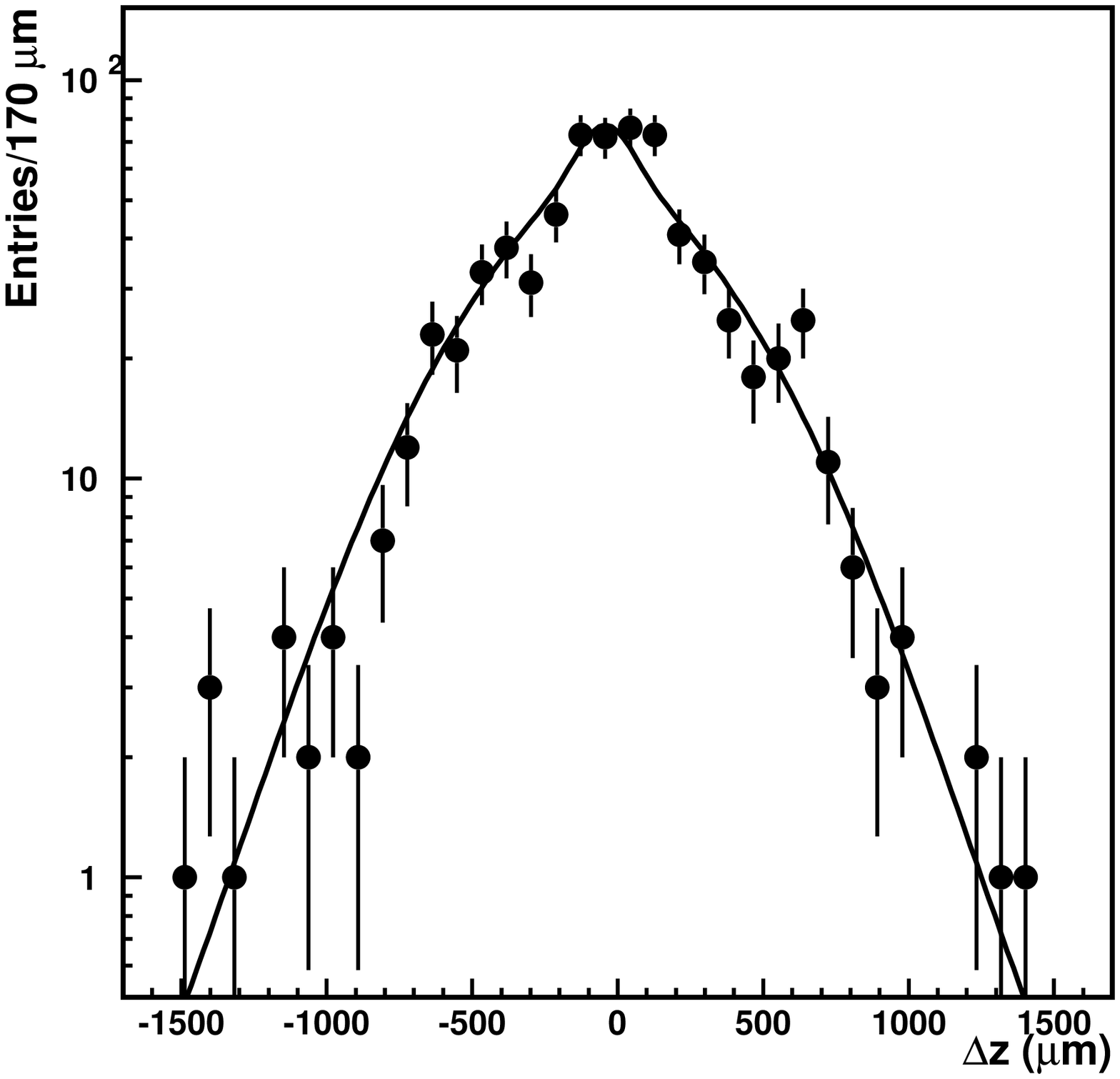} \\
    \vskip -60mm \hskip 45mm (b) \\ \vskip 56mm
\caption{Semilogarithmic plot of the $\Delta z$ distributions from
MC simulation for peaked backgrounds for (a) opposite-flavor and
(b) same-flavor events.} \label{fig:bkgpeaked}
\end{center}
\end{figure}

The parameterizations of all background shapes
are given in Eqs.~(\ref{eqn:bkgpdf1}) and~(\ref{eqn:bkgpdf2}).
Figs.~\ref{fig:bkgunpeak} and~\ref{fig:bkgpeaked} show the $\Delta
z$ distributions for unpeaked and peaked backgrounds,
respectively. The unbinned maximum likelihood fitting method is
applied to extract the background parameters.

We use the $\Delta z$ distribution from the $D^0$ missing mass
sideband to reproduce the unpeaked background shape.

For the $\Delta z$ shapes of the peaked background, we use the
Monte Carlo simulation. We find that the probability given by
Eq.~(\ref{eqn:pkbkgof}) for the opposite flavor sample gives a
good fit even without including the mixing term. The fraction of
mixing, $f_1$, is then set to zero and fixed for the fit to the
opposite flavor sample. Therefore, the floating parameters are
$f_0$ and $\tau_{\rm bkg}$. For the same flavor events, inclusion
of the mixing term is necessary to obtain a good fit and the
floating parameters are $f_0$, $f_1$ and $\tau_{\rm bkg}$. The
mixing frequency in the background is determined from the fit.

\section{FITTING RESULTS}
\label{sec:fit_results}

We now discuss the result of the fit. In the final fit, $\Delta
m_d$ is the only free parameter. The parameters of the resolution
functions and background shapes are fixed to their central values.
The uncertainties in these parameters are included in the
systematic error. The parameters used in the fit are the
following:
\begin{itemize}
\item The $B^0$ lifetime, $\tau_{B^0} = (1.548 \pm 0.032) \rm ~ps$, is fixed
to the PDG2000 value~\cite{PDG2000}.

\item The signal and unpeaked background resolution function are determined from
$J/\psi \to l^+ l^-$ decays in data. The parameters of
Eq.~(\ref{eqn:reso}) are: $\rm \sigma_1 = (94 \pm 6)~\mu m$, $\rm
\sigma_2 = (227\pm 18)~\mu m$, $\rm \sigma_3 = (736 \pm 98)~\mu m$,
$f_1 = 0.56 \pm 0.07$ and $f_2 = 0.38 \pm 0.04$. In the
fit, all the parameters are fixed to their central values.

\item The peaked background resolution function is determined from $J/\psi
\to l^+ l^-$ decays in MC convolved with an additional 50 micron smearing
term. The parameters of Eq.~(\ref{eqn:reso}) are: $\rm \sigma_1 =
(93 \pm 4)~\mu m$, $\rm \sigma_2 = (209\pm 13)~\mu m$, $\rm
\sigma_3 = (627 \pm 74)~\mu m$, $f_1 = 0.49 \pm 0.08$ and $f_2 =
0.43 \pm 0.04$. In the fit, all the parameters are fixed to their
central values. Note that the offset of the second Gaussian is
$\rm \mu_2 = (81 \pm 45) ~\mu m$ for the same flavor peaked
background.

\item The parameters of the unpeaked background shapes are fixed to values
determined from fitting the $D^0$ missing mass sideband (see
Fig.~\ref{fig:bkgunpeak}). The parameters for opposite flavor
events are: $f_0 = 0.37 \pm 0.06, \rm ~\tau_{bkg} = (1.98 \pm
0.17) ~ps$. For same flavor events: $f_0 = 0.25 \pm 0.06, \rm
~\tau_{bkg} = (1.41 \pm 0.10) ~ps$.

\item The parameters of the peaked background shape are fixed to values
determined from fitting the Monte Carlo simulation in
Fig.~\ref{fig:bkgpeaked}. Here, $\Delta m_d$ is fixed to $0.509
~{\rm ps}^{-1}$. The parameters for opposite flavor events are:
$f_0 = 0.07 \pm 0.07, \rm ~\tau_{bkg} = (1.34 \pm 0.12) ~ps$. The
fit gives $f_1 = 0$. For same flavor events: $f_0 = 0.10 \pm
0.04,~f_1 = 0.40 \pm 0.05,\rm ~\tau_{bkg} = (1.84 \pm 0.26) ~ps$.

\item The fractions $f_{\rm bkg} = 0.30 \pm 0.01$, $f_{\rm bkg}^{\rm OF} = 0.69 \pm 0.06$, $f_{\rm bkg}^{\rm
OF-unpeak}= 0.49 \pm 0.03$ and $f_{\rm bkg}^{\rm SF-unpeak} = 0.53
\pm 0.04$ are determined by fitting the $D^0$ missing mass
distribution in Fig.~\ref{fig:mmdosss}.

\end{itemize}

\begin{figure}[tb]
\begin{center}
    \includegraphics[height=7cm]{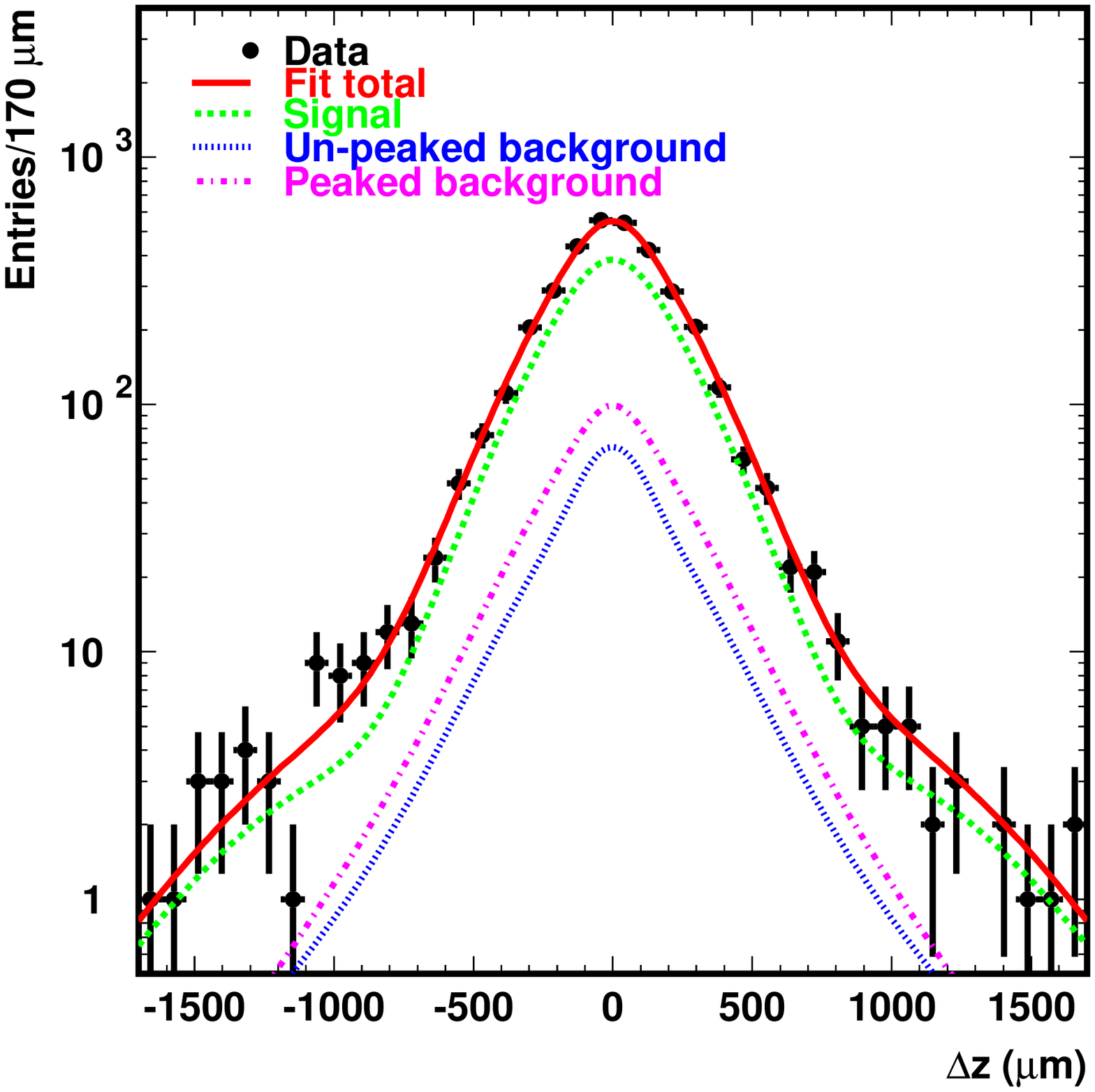}\\
    \vskip -61mm \hskip 45mm (a) \\ \vskip 51mm
    \includegraphics[height=7cm]{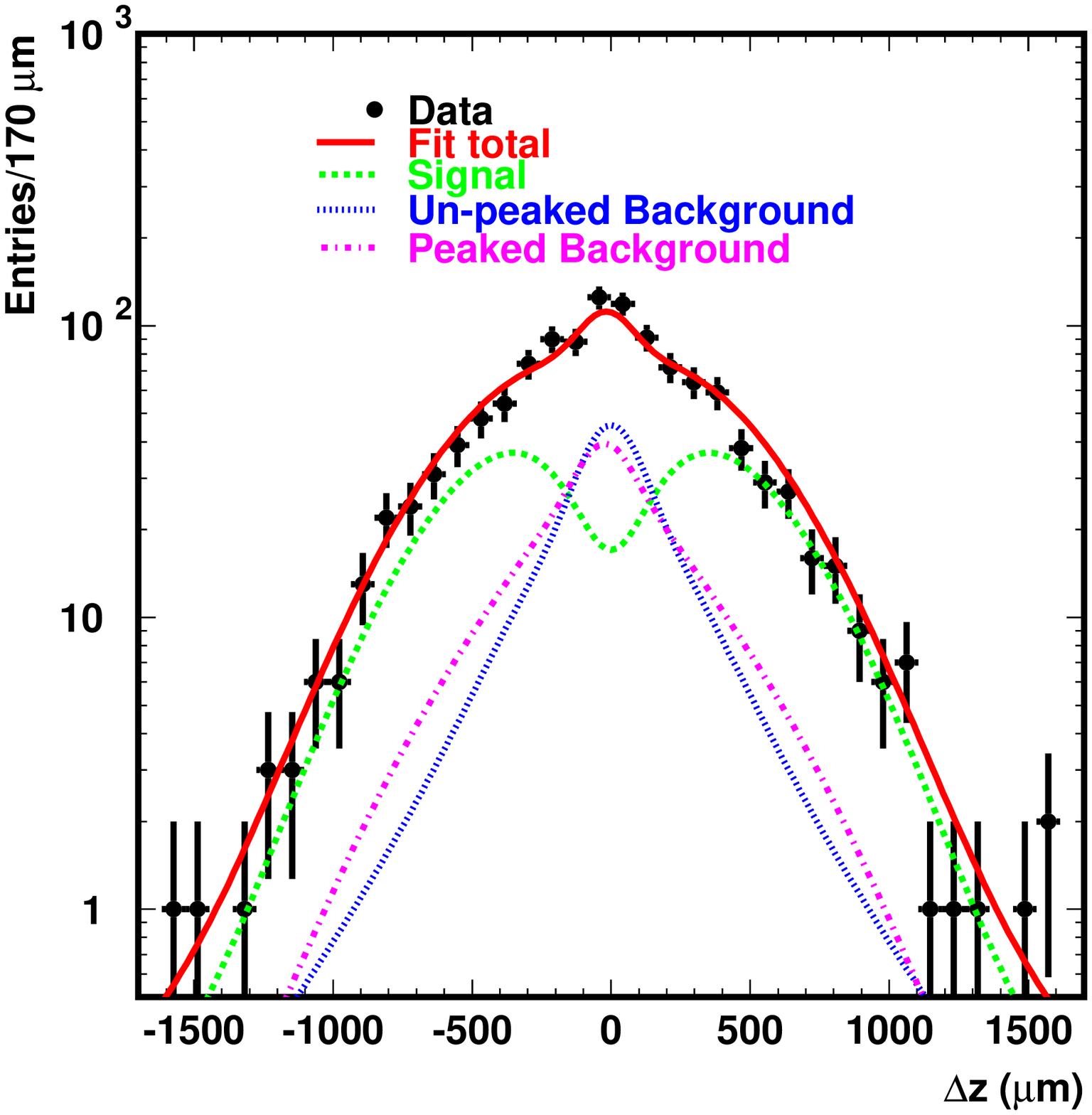} \\
    \vskip -61mm \hskip 45mm (b) \\ \vskip 57mm
\caption{$\Delta z$ distributions for (a) opposite-flavor and (b)
same-flavor events in data. The curve from the fit and the
contributions of signal, unpeaked and peaked backgrounds are
overlaid.} \label{fig:dmfit}
\end{center}
\end{figure}

\begin{figure}[tb]
\begin{center}
    \includegraphics[height=7cm]{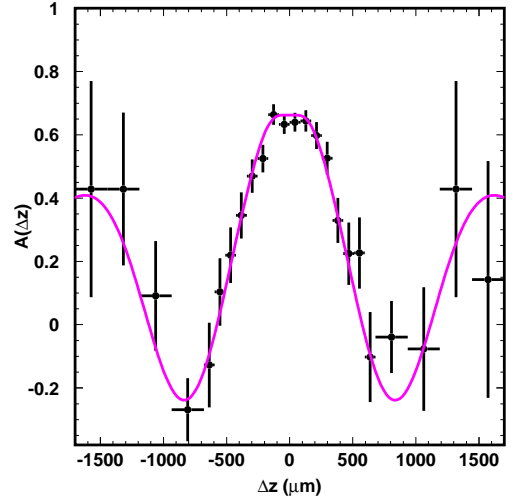}\\
\caption{Distribution of the asymmetry, $A(\Delta z)$, as a
function of $\Delta z$ for data with the fit curve overlaid.}
\label{fig:osclik}
\end{center}
\end{figure}

Using the parameters determined above, $\Delta m_d$ is extracted
from the fit. We find

\begin{eqnarray}
\Delta m_d = (0.509 \pm 0.017~({\rm stat}))~{\rm ps^{-1}}.
\end{eqnarray}
In Fig.~\ref{fig:dmfit} we show the $\Delta z$ distributions for
SF and OF data together with the curves from the fit. To display
the charge asymmetry (see Fig.~\ref{fig:osclik}) between SF and OF
events, we use
\begin{eqnarray}
\label{eqn:chargeaysm} A(\Delta z) & \equiv & {{\rm N^{OF}}(\Delta
z)-{\rm N^{SF}}(\Delta z) \over {\rm N^{OF}}(\Delta z)+{\rm
N^{SF}}(\Delta z)} ~,
\end{eqnarray}
where ${\rm N}(\Delta z)$ is the yield of the signal candidates as
a function of $\Delta z$.

\section{VALIDATION CHECK}
\label{sec:vcheck} 
We also perform a series of fits to signal
Monte Carlo using the same procedure used in the analysis of the
data. We generate four signal MC samples with different $\Delta
m_d$ values. Table~\ref{tab:sigmcchk} summarizes the fitting
results and shows the consistency between the input and extracted
$\Delta m_d$. We conclude there is no significant fit bias.

\begin{table}[!htb]
\begin{center}
\caption{Summary of signal MC bias test.}
\label{tab:sigmcchk}
\begin{ruledtabular}
\begin{tabular}{cc}
$\Delta m_d~({\rm ps^{-1}})$ for MC generation & $\Delta m_d \rm ~(ps^{-1})$ from fit yields \\
\hline
0.442 & $0.448 \pm 0.011$ \\
0.472 & $0.470 \pm 0.012$ \\
0.502 & $0.492 \pm 0.011$ \\
0.532 & $0.537 \pm 0.012$ \\
\end{tabular}
\end{ruledtabular}
\end{center}
\end{table}

\begin{figure*}[tb]
\begin{center}
    \includegraphics[height=8cm]{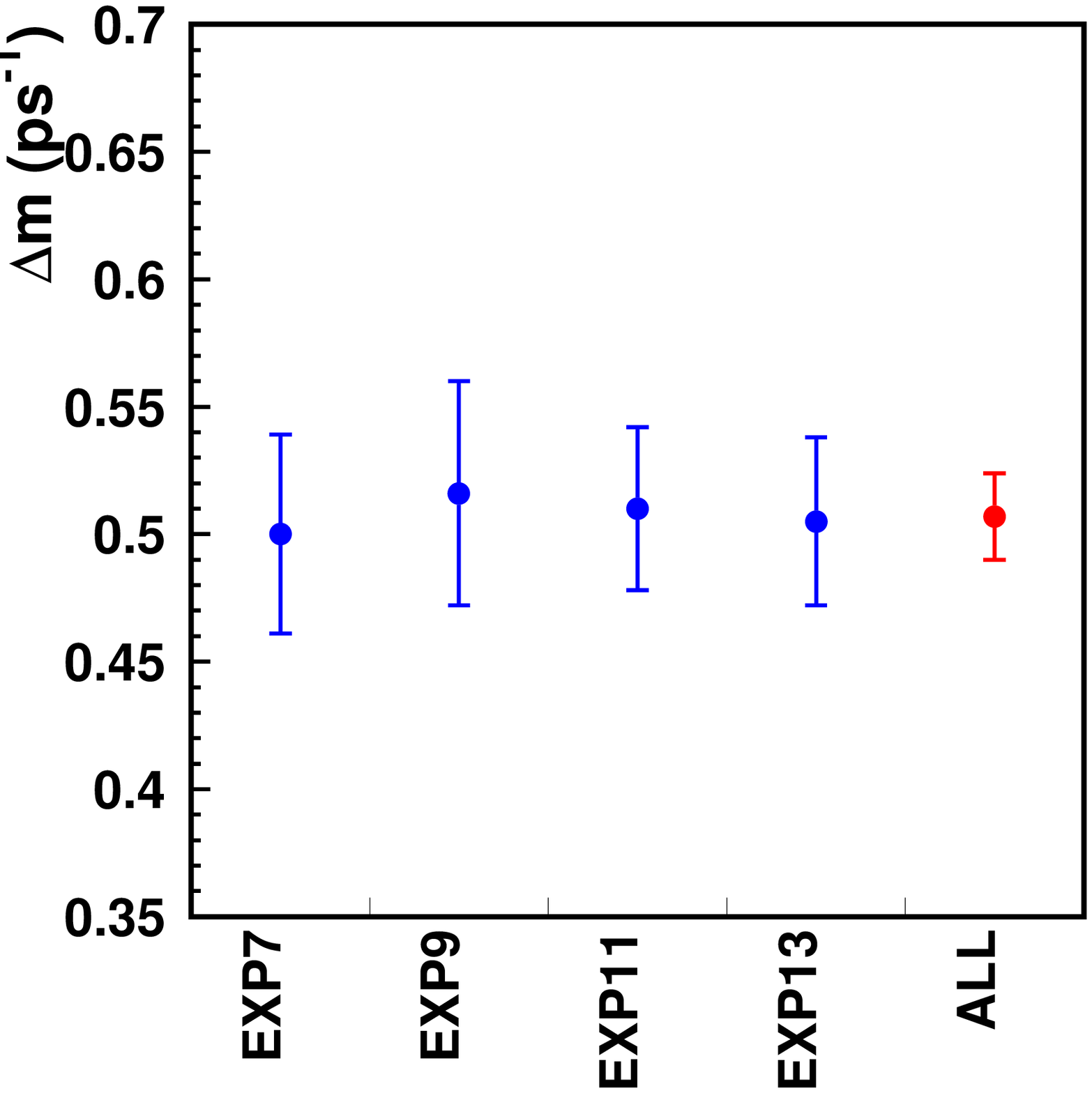}
    \includegraphics[height=8cm]{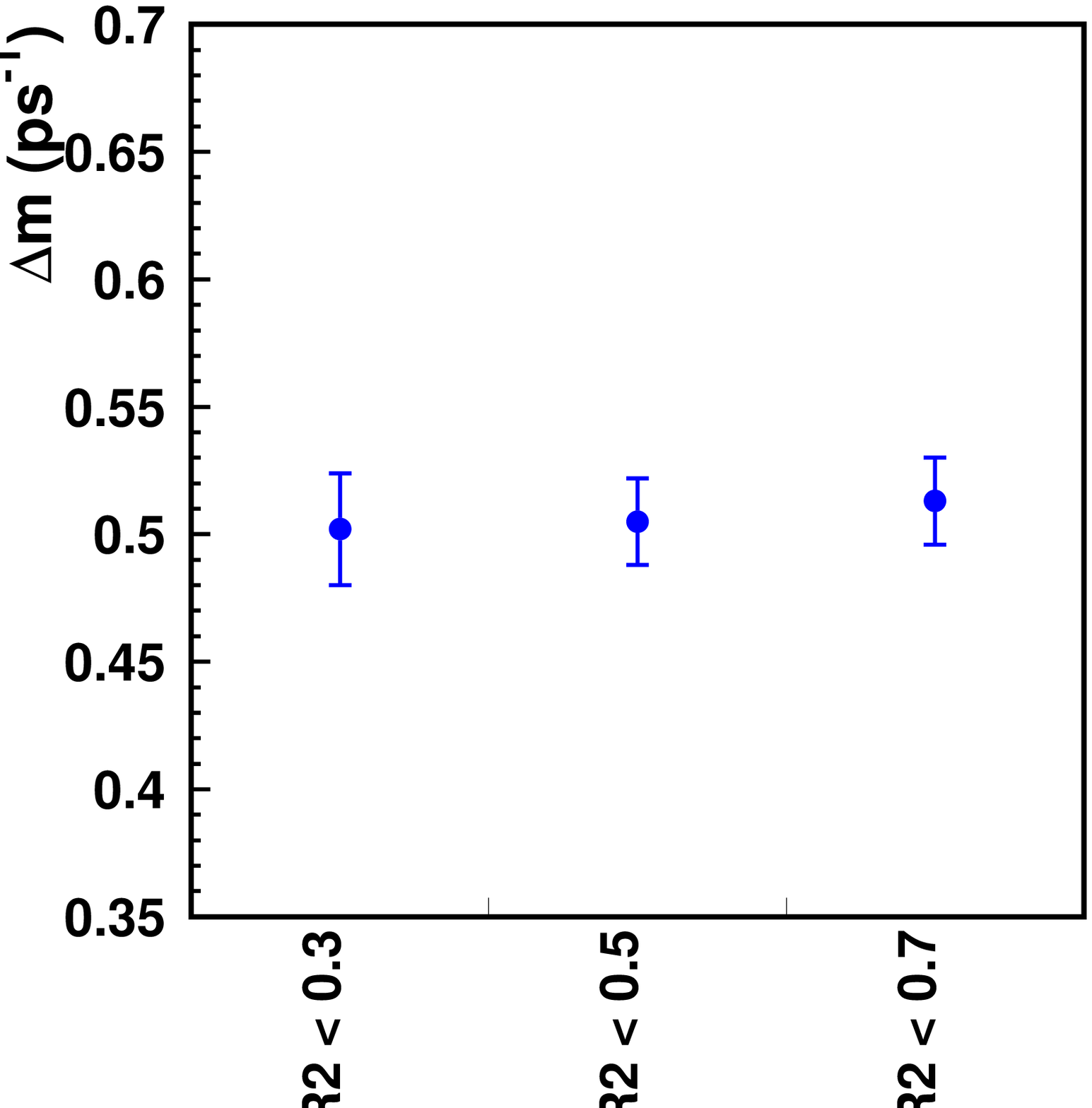} \\
    \vskip -68mm \hskip -47mm (a) \hskip 75mm (b) \\ \vskip 60mm
    \includegraphics[height=8cm]{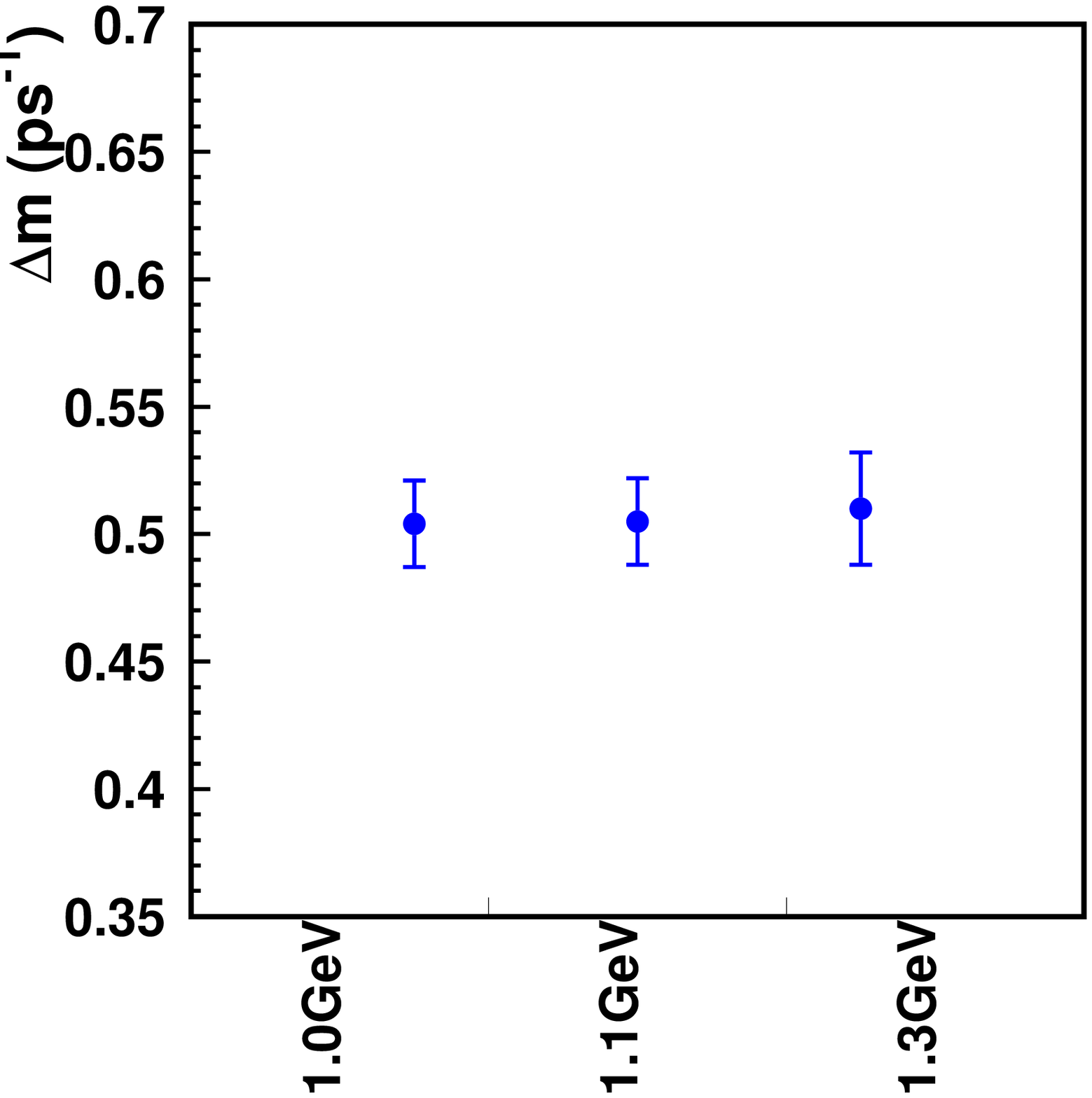}
    \includegraphics[height=8cm]{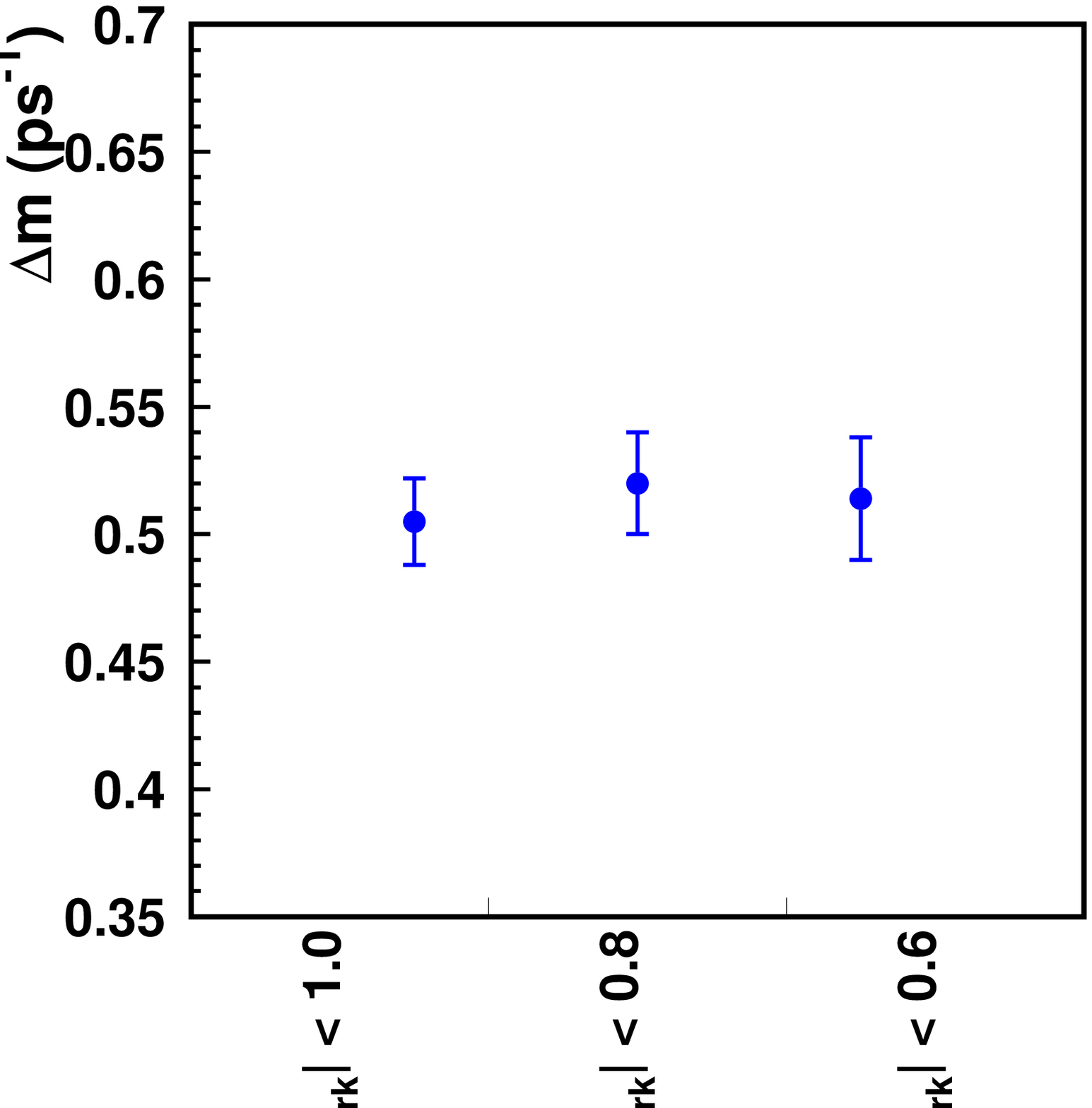} \\
    \vskip -68mm \hskip -47mm (c) \hskip 75mm (d) \\ \vskip 73mm
\caption{$\Delta m_d$ as function of (a) experiment number, (b)
$R_2$ cut value, (c) cut on the lepton momentum in the CM frame,
(d) lepton fiducial angle requirement.} \label{fig:dmvs}
\end{center}
\end{figure*}

We also check for run dependence of the $\Delta m_d$ extraction.
The value of $\Delta m_d$ is shown for four different experimental
data taking periods (referred to as experiments 7, 9, 11, 13). The
results are shown in Fig.~\ref{fig:dmvs}(a).

To check the dependence of the $\Delta m_d$ fitting on the
continuum background level, we also calculate $\Delta m_d$ as a
function of the cut on the normalized second Fox-Wolfram
moment~\cite{foxwolfram} $R_2=H_2/H_0$. The results are shown in
Fig.~\ref{fig:dmvs}(b).

We have also examined the variation of the fit results as the
values of the cuts on the lepton momentum $p^*_l$ in the CM frame
and the fiducial angle requirement for the lepton and $\pi_f$. The
results are shown in Figs~\ref{fig:dmvs}(c), (d). We find that all
the variations are consistent with statistical fluctuations.

Furthermore, we performed fits separately for SF and OF events.
Table~\ref{tab:ssos} shows the fit results for SF events and OF
events, respectively.
\begin{table}[!htb]
\begin{center}
\caption{Summary of $\Delta m_d$ fit results for SF (OF) events. Only
statistical errors are shown.}
\label{tab:ssos}
\begin{ruledtabular}
\begin{tabular}{lcc}
Sample & Events & $\Delta m_d \rm ~(ps^{-1})$\\
\hline
SF & 1213 & $ \Delta m_d = 0.511 \pm 0.035 $ \\
OF & 3686 & $ \Delta m_d = 0.504 \pm 0.019 $ \\
\end{tabular}
\end{ruledtabular}
\end{center}
\end{table}

We also fit $e$ tagged events and $\mu$ tagged events separately (see
Table~\ref{tab:emu}).
\begin{table}[!htb]
\begin{center}
\caption{Summary of $\Delta m_d$ fit results for $e$ and $\mu$ tagged
events. Only statistical errors are shown.}
\label{tab:emu}
\begin{ruledtabular}
\begin{tabular}{lcc}
Sample & Events & $\Delta m_d \rm ~(ps^{-1})$\\
\hline
$e$ tagged & 2324 & $ \Delta m_d = 0.510 \pm 0.025 $ \\
$\mu$ tagged & 2575 & $ \Delta m_d = 0.503 \pm 0.024 $ \\
\end{tabular}
\end{ruledtabular}
\end{center}
\end{table}

To check the sensitivity of the result to tails of the vertex
resolution function, we vary the $\Delta z$ range of the fit. The
results are shown in Table~\ref{tab:dzwin} and are consistent with
the primary result.
\begin{table}[!htb]
\begin{center}
\caption{Summary of $\Delta m_d$ fit results for different $\Delta z$
window. Only statistical errors are shown.}
\label{tab:dzwin}
\begin{ruledtabular}
\begin{tabular}{cc}
$\max(|\Delta z|)~(\rm \mu m)$ & $\Delta m_d \rm ~(ps^{-1})$\\
\hline
1700 & $ 0.509 \pm 0.017 $ \\
1275 & $ 0.507 \pm 0.017 $ \\
 875 & $ 0.511 \pm 0.019 $ \\
 425 & $ 0.520 \pm 0.024 $ \\
\end{tabular}
\end{ruledtabular}
\end{center}
\end{table}

\section{Systematic Uncertainties} 
\label{sec:syserr}

Various possible sources of
systematic uncertainties were investigated. In
Table~\ref{tab:syserr}, we list the contributions to the
systematic errors in $\Delta m_d$ that were estimated by varying
the relevant parameters by one standard deviation.
\begin{table}[!htb]
\begin{center}
\caption{Contributions to the systematic error.}
\label{tab:syserr}
\begin{ruledtabular}
\begin{tabular}{lc}
Source & Errors $ \rm (ps^{-1})$\\
\hline
Background fraction & $\pm 0.014$ \\
Signal resolution function & $\pm 0.012$ \\
Background shape & $\pm 0.005$ \\
$B^0$ lifetime $\rm (1.548\pm 0.032~ps)$ & $\pm 0.005$\\
Detector resolution $\rm (50\pm 18~\mu m)$ & $\pm 0.002$\\
\hline
Total & $\pm 0.020$\\
\end{tabular}
\end{ruledtabular}
\end{center}
\end{table}

\begin{itemize}
\item $\mathbf{Background~fraction}$ \\
As described in Section~\ref{sec:pdflike}, the background parts of
the likelihood function are weighted by $f_{\rm bkg}$, $f_{\rm
bkg}^{OF}$, $f_{\rm bkg}^{OF-unpeak}$ and $f_{\rm
bkg}^{SF-unpeak}$ in Eq.~(\ref{eqn:likli}), which are determined
from a fit to the $D^0$ missing mass distribution in data. The
corresponding systematic error, $\pm 0.014$, is estimated by
varying the parameters of these background fractions by $\pm 1
\sigma$.
\item $\mathbf{Signal~resolution~function}$ \\
The fitted parameters of the $\Delta z$ distributions for $J/\psi
\to l^+ l^-$ events are varied according to the limited data
statistics by $\pm 1 \sigma$. The difference from the standard fit
is taken as a measure of systematic error for each parameter. The
corresponding uncertainty for each parameter is added in
quadrature in order to estimate a systematic uncertainty in
$\Delta m_d$ of $\pm 0.012 \rm ~ps^{-1}$.
\item $\mathbf{Background~shape}$ \\
For both peaked and unpeaked backgrounds, the $\Delta z$
distributions are determined from a method described in
Section~\ref{sec:bkgdis}. Varying all parameters by $\pm 1
\sigma$, we take the differences, $\pm 0.004$, from the standard
fit as the systematic error.
\item $B^0$ $\mathbf{lifetime}$ \\
The value of the $B^0$ lifetime was varied according to the
uncertainty of the PDG2000~\cite{PDG2000} value. Changes in
$\Delta m_d$ of $\pm 0.005$ are observed.
\item $\mathbf{Detector~resolution}$ \\
In Section~\ref{sec:bkgdis}, we mentioned that the MC background
distributions are smeared to correct for the difference in the
vertex resolution between the MC prediction and the data. The
smearing function was a Gaussian with $\sigma = 50~\mu$m. We
varied the amount of smearing by its uncertainty
($18~\mu$m)~\cite{christo} and repeated the fit. We obtain a
difference in $\Delta m_d$ of $\pm 0.002$
\end{itemize}
The total systematic error, obtained by summing all errors from
the different sources in quadrature, is $\pm 0.020 \rm ~ps^{-1}$.

\section{CONCLUSION}
\label{sec:conclusion} Using 29.1 $\rm fb^{-1}$ of data collected
with the Belle detector at the $\Upsilon(4S)$, we have measured
the $B^0 - \overline{B}^0$ mixing frequency $\Delta m_d$ in
$B^0(\overline{B}^0) \to D^{*\mp} \pi^{\pm}$ decay with a partial
reconstruction technique. The data were accumulated between
January 2000 and July 2001.

The asymmetric $e^+ e^-$ beam energies of the KEKB collider allows
for the extraction of the time evolution of the $B$ meson wave
function from precise measurements of the decay vertex positions.
The data are separated into SF and OF samples. The $B^0$ decay
vertex resolution for the signal was determined from the $\Delta
z$ distribution of $J/\psi \to l^+ l^-$ decays that occur in the
same data sample. The backgrounds are divided into two components,
peaked and unpeaked backgrounds. The $\Delta m_d$ value, obtained
by simultaneously fitting the SF and OF time distributions, is
\begin{eqnarray*}
\Delta m_d &=& (0.509 \pm 0.017(\rm stat) \pm 0.020 (\rm
syst))~ps^{-1}.
\end{eqnarray*}
This is the first direct measurement of $\Delta m_d$ using the
technique of partial reconstruction of $B^0(\overline{B}^0) \to
D^{*\mp} \pi^{\pm}$ decays.  This measurement is statistically
uncorrelated with all other reported experimental results for
$\Delta m_d$.  It is also almost systematically independent from
all other measurements. The systematic uncertainty is dominated by
the background fractions and the signal resolution function; all
quantities are measured experimentally except for the peaked
background. This measurement agrees with the world average value
of $\Delta m_d = (0.472 \pm 0.017) \rm ~ps^{-1}$~\cite{PDG2000},
and serves as a validation of the technique that will be used in
the future for the measurement of the $CP$ violation parameter
$\sin(2 \phi_1 + \phi_3)$~\cite{zyh}.

\vskip 3mm
We wish to thank the KEKB accelerator group for the excellent
operation of the KEKB accelerator.
We acknowledge support from the Ministry of Education,
Culture, Sports, Science, and Technology of Japan
and the Japan Society for the Promotion of Science;
the Australian Research Council
and the Australian Department of Industry, Science and Resources;
the National Science Foundation of China under contract No.~10175071;
the Department of Science and Technology of India;
the BK21 program of the Ministry of Education of Korea
and the CHEP SRC program of the Korea Science and Engineering Foundation;
the Polish State Committee for Scientific Research
under contract No.~2P03B 17017;
the Ministry of Science and Technology of the Russian Federation;
the Ministry of Education, Science and Sport of Slovenia;
the National Science Council and the Ministry of Education of Taiwan;
and the U.S.\ Department of Energy.

\bibliographystyle{unsrt}
\bibliography{mix_bib}

\begin{thebibliography}{10}

\bibitem{Sandabook}
I.~I. Bigi and A.~I. Sanda.
\newblock {\em ``$CP$ Violation''}.
\newblock Cambridge University Press, Cambridge, 2000.

\bibitem{buras1}
A. J. Buras {\it et al.}, Nucl. Phys. B {\bf 347}, 491 (1990).

\bibitem{Inamiburas2}
T. Inami and C. S. Lim, Prog. Theor. Phys. {\bf 65}, 297 and 1772 (1981); A. J.
  Buras, Phys. Rev. Lett. {\bf 46}, 1354 (1981).

\bibitem{donoghue}
J. F. Donoghue, E. Golowich and B. R. Holstein. {\it ``Dynamics of the Standard
  Model''}. Cambridge University Press, Cambridge, 1992.

\bibitem{dunietz}
I. Dunietz and J. L. Rosner, Phys. Rev. D {\bf 34}, 1404 (1986); I. Dunietz,
  Phys. Lett. B {\bf 427}, 179-182 (1998).

\bibitem{KEKB}
S.~Kurokawa {\it et al.}, KEK Preprint 2001-157 (2001), to appear in Nucl.
  Instr. and Meth. A.

\bibitem{Belle}
A.~Abashian {\it et al.} (Belle Collaboration), Nucl. Instr. and Meth. {\bf
  A479}, 117 (2002).

\bibitem{SVD}
G.~Alimonti {\it et al.}, Nucl. Instr. and Meth. {\bf A453}, 71 (2000).

\bibitem{CDC}
H.~Hirano {\it et al.}, Nucl. Instr. and Meth. {\bf A455}, 294 (2000);
  M.~Akatsu {\it et al.}, Nucl. Instr. and Meth. {\bf A454}, 322 (2000).

\bibitem{TOF}
H.~Kichimi {\it et al.}, Nucl. Instr. and Meth. {\bf A453}, 315 (2000).

\bibitem{ACC}
T.~Iijima {\it et al.}, Nucl. Instr. and Meth. {\bf A453}, 321 (2000).

\bibitem{ECL}
H.~Ikeda {\it et al.}, Nucl. Instr. and Meth. {\bf A441}, 401 (2000).

\bibitem{EID}
K.~Hanagaki {\it et al.}, submitted to Nucl. Instr. and Meth., hep-ex/0108044.

\bibitem{KLM}
A.Abashian {\it et al.}, Nucl. Instr. and Meth. {\bf A449}, 112 (2000).

\bibitem{QQ}
The QQ $B$ meson decay event generator was developed by
  the~CLEO~Collaboration.~See the following URL:\\
  http://www.lns.cornell.edu/public/CLEO/soft/QQ.

\bibitem{GEANT}
CERN Program Library Long Writeup W5013, CERN, 1993.

\bibitem{hadsel}
K. Abe {\it et al.} (Belle Collaboration), Phys. Rev. D {\bf 64}, 072001
  (2001).

\bibitem{partial}
R. Giles {\it et al.} (CLEO Collaboration), Phys. Rev. D {\bf 30}, 2279 (1984);
  G. Brandenburg {\it et al.} (CLEO Collaboration), Phys. Rev. Lett. {\bf 80},
  2762 (1998); B. H. Behrens {\it et al.} (CLEO Collaboration), Phys. Lett. B
  {\bf 490}, 36-44 (2000).

\bibitem{PDG2000}
D.E.~Groom~{\it et al.} Particle Data~Group, Eur. Phys. J. {\bf C15}, 1 (2000).

\bibitem{zyh}
Yangheng Zheng, PhD dissertation, University of Hawaii at Manoa, Department of
  Physics, 2002.

\bibitem{drho}
M. S. Alam {\it et al.} (CLEO Collaboration), Phys. Rev. D {\bf 50}, 43-68
  (1994).

\bibitem{cystalballfunc}
T. Skwarnicki, PhD dissertation, Institute for Nuclear Physics, Krakow 1986;
  DESY Internal Report, DESY F31-86-02 (1986).

\bibitem{foxwolfram}
G.~Fox and S.~Wolfram, Phys. Rev. Lett {\bf 41}, 1581 (1978).

\bibitem{christo}
Christos Leonidopoulos, PhD dissertation, Princeton University, Department of
  Physics, 2000.

\end{thebibliography}

\appendix
\section{Background PDFs}
\label{sec:app:bpdfs} For the background, the PDFs of OF and SF
events are parameterized by
\begin{eqnarray}
\label{eqn:bkgpdf1} \nonumber F^{\rm OF}_{\rm bkg}(\Delta t) &=&
\int^{\infty}_{-\infty} (f_{\rm bkg}^{\rm OF-unpeak}{\rm P}_{\rm
bkg}^{\rm OF-unpeak}(\Delta t') \\
\nonumber &&~~~~~\times R_{\rm bkg}^{\rm unpeak}(\Delta t - \Delta t') \\
\nonumber &&~~~+ (1 - f_{\rm bkg}^{\rm OF-unpeak}){\rm P}_{\rm
bkg}^{\rm
OF-peak}(\Delta t') \\
&&~~~~~\times R_{\rm bkg}^{\rm peak}(\Delta t - \Delta t') )
 d \Delta t'~,\\
\label{eqn:bkgpdf2} \nonumber F^{\rm SF}_{\rm bkg}(\Delta t) &=&
\int^{\infty}_{-\infty} (f_{\rm bkg}^{\rm SF-unpeak}{\rm P}_{\rm
bkg}^{\rm SF-unpeak}(\Delta t') \\
\nonumber &&~~~~~\times R_{\rm bkg}^{\rm unpeak}(\Delta t - \Delta t') \\
\nonumber &&~~~+ (1 - f_{\rm bkg}^{\rm SF-unpeak}){\rm P}_{\rm
bkg}^{\rm
SF-peak}(\Delta t') \\
&&~~~~~\times R_{\rm bkg}^{\rm peak}(\Delta t - \Delta t') )
 d \Delta t'~.
\end{eqnarray}
Here, $f_{\rm bkg}^{\rm SF-unpeak} \equiv N_{\rm bkg}^{\rm
SF-unpeak} / N_{\rm bkg}^{\rm SF}$ and $f_{\rm bkg}^{\rm
OF-unpeak} \equiv N_{\rm bkg}^{\rm OF-unpeak} / N_{\rm bkg}^{\rm
OF}$ are the fractions of unpeaked background in the SF and OF
events. $N$ is the number of events, the subscripts $\rm bkg$ and
$\rm sig$ denote the background and signal components. $R_{\rm
bkg}^{\rm peak}(\Delta t)$ and $R_{\rm bkg}^{\rm unpeak}(\Delta
t)$ are resolution functions for peaked and unpeaked background
and are parameterized by triple-Gaussian distributions, which have
the same form as the signal resolution function $R_{\rm
sig}(\Delta t)$ in Eq.~(\ref{eqn:reso}). In the fit, we use the
signal resolution function for $R_{\rm bkg}^{\rm unpeak}(\Delta
t)$. The peaked background resolution function $R_{\rm bkg}^{\rm
peak}(\Delta t)$, described in Section~\ref{sec:reso}, is
determined from $J/\psi \to l^+ l^-$ decays in MC with additional
smearing.

In the unpeaked background PDF, ${\rm P}_{\rm bkg}^{\rm
OF-unpeak}(\Delta t)$ and ${\rm P}_{\rm bkg}^{\rm
SF-unpeak}(\Delta t)$ are defined by
\begin{eqnarray}
\label{eqn:upkbkgof} \nonumber {\rm P}_{\rm bkg }^{\rm
OF-unpeak}(\Delta
t)&=&f_0^{\rm OF-unpeak}\delta (\Delta t) \\
&& \hskip -5mm +(1 - f_0^{\rm OF-unpeak}){1 \over {2 \tau_{\rm
bkg}}}{e^{-{|\Delta t| \over
\tau_{\rm bkg}}}}~, \\
\label{eqn:upkbkgsf} \nonumber {\rm P}_{\rm bkg}^{\rm
SF-unpeak}(\Delta t)
&=& f_0^{\rm SF-unpeak}\delta (\Delta t) \\
&& \hskip -5mm + (1 - f_0^{\rm SF-unpeak}){ 1 \over {2 \tau_{\rm
bkg}} }{e^{-{|\Delta t| \over \tau_{\rm bkg}}}}~.
\end{eqnarray}

In the peaked background PDF, ${\rm P}_{\rm bkg}^{\rm
OF-peak}(\Delta t)$ and ${\rm P}_{\rm bkg}^{\rm SF-peak}(\Delta
t)$ are defined by
\begin{eqnarray}
\label{eqn:pkbkgof} \nonumber {\rm P}_{\rm bkg }^{\rm
OF-peak}(\Delta t)&=& f_0^{\rm OF-peak} \delta (\Delta t) \\
\nonumber &&\hskip -14mm + (1 - f_0^{\rm OF-peak})(1-f_1^{\rm
OF-peak}){1 \over {2
\tau_{\rm bkg}}}{e^{-{|\Delta t| \over \tau_{\rm bkg}}}} \\
\nonumber &&\hskip -14mm + (1 - f_0^{\rm OF-peak})f_1^{\rm
OF-peak}{\frac{1 + \Delta m_d^2 \tau_{\rm bkg}^2}{4 \tau_{\rm bkg}
+ 2 \Delta m_d^2
\tau_{\rm bkg}^3}} \\
&&\hskip -8mm \times~ {e^{-{|\Delta
t| \over \tau_{\rm bkg}}}\left[1+\cos(\Delta m_d \Delta t)\right]}~,\\
\label{eqn:pkbkgsf} \nonumber {\rm P}_{\rm bkg }^{\rm
SF-peak}(\Delta t)&=&f_0^{\rm SF-peak} \delta (\Delta t) \\
\nonumber && \hskip -14mm + (1 - f_0^{\rm SF-peak})(1-f_1^{\rm
SF-peak}){1 \over {2
\tau_{\rm bkg}}}{e^{-{|\Delta t| \over \tau_{\rm bkg}}}} \\
\nonumber &&\hskip -14mm + (1 - f_0^{\rm SF-peak})f_1^{\rm
SF-peak}{\frac{1 + \Delta m_d^2
\tau_{\rm bkg}^2}{2 \Delta m_d^2 \tau_{\rm bkg}^3}}\\
&&\hskip -8mm \times~{e^{-{|\Delta t| \over \tau_{\rm bkg}}}
\left[1-\cos(\Delta m_d \Delta t)\right]}~.
\end{eqnarray}
In the above equations, $f_0$ is the prompt lifetime fraction and
$(1 - f_0 ) f_1$ is the mixing fraction. $\tau_{\rm bkg}$ is the
lifetime of the background component that does not originate from
a prompt source. $\Delta m_d$ contributes to the peaked background
since the peaked background is dominated by $B^0$
decays~\cite{zyh}. Thus, mixing terms are included in the PDFs for
the peaked background. In the unpeaked background expression, we
set the mixing term to zero since a parameterization without
mixing reproduces the data well.

\end{document}